\documentclass{article}

\usepackage{fancybox}
\usepackage{pstricks}
\usepackage{amsfonts}
\usepackage[bookmarks=true,
            bookmarksnumbered=true,
            bookmarksopen=false,
            plainpages=false,
            pdfpagelabels,
            colorlinks,
            citecolor=black,              
            pagecolor=black,         
            linkcolor=black,         
            menucolor=black,         
            urlcolor=black,       
            linkcolor=black]{hyperref}
\usepackage{graphicx}
\usepackage{algorithm}
\usepackage{algorithmic}
\usepackage{amsmath}
\usepackage{amssymb}
\usepackage[square,comma,sort]{natbib}
\usepackage{authblk}

\pdfoutput=1
\relax
\title{A Class of Parallel Tiled Linear Algebra Algorithms for Multicore Architectures}

\author[1]{Alfredo Buttari}
\author[2]{Julien Langou}
\author[1]{Jakub Kurzak}
\author[1,3,4]{Jack Dongarra}

\affil[1]{Department of Electrical Engineering and Computer
 Science, University Tennessee, Knoxville, Tennessee}
\affil[2]{Department of Mathematical and Statistical Sciences,
University of Colorado Denver,
Colorado}
\affil[3]{Oak Ridge National Laboratory, Oak
Ridge, Tennessee}
\affil[4]{University of Manchester, Manchester UK}

\begin{document}
\maketitle
\begin{abstract}
  As multicore systems continue to gain ground in the High Performance
  Computing world, linear algebra algorithms have to be reformulated
  or new algorithms have to be developed in order to take advantage of
  the architectural features on these new processors. Fine grain
  parallelism becomes a major requirement and introduces the necessity
  of loose synchronization in the parallel execution of an operation.
  This paper presents algorithms for the Cholesky, LU and QR
  factorization where the operations can be represented as a sequence
  of small tasks that operate on square blocks of data. These tasks
  can be dynamically scheduled for execution based on the dependencies
  among them and on the availability of computational resources. This
  may result in an out of order execution of the tasks which will
  completely hide the presence of intrinsically sequential tasks in
  the factorization.  Performance comparisons are presented with the
  LAPACK algorithms where parallelism can only be exploited at the
  level of the BLAS operations and vendor implementations. The
  described approach shows encouraging results, continuing the trend
  established in previous work by the same
  authors~\cite{cell_chol,tiledqr,Kurzak:2006:ILA}.  
\end{abstract}

\section{Introduction}
\label{sec:intro}
In the last twenty years, microprocessor manufacturers have been
driven towards higher performance rates only by the exploitation of
higher degrees of {\em Instruction Level Parallelism} (ILP). Based on
this approach, several generations of processors have been built where
clock frequencies were higher and higher and pipelines were deeper and
deeper. As a result, applications could benefit from these innovations
and achieve higher performance simply by relying on compilers that
could efficiently exploit ILP. Due to a number of physical limitations
(mostly power consumption and heat dissipation) this approach cannot
be pushed any further. For this reason, chip designers have moved
their focus from ILP to {\em Thread Level Parallelism} (TLP) where
higher performance can be achieved by replicating execution units (or
{\em cores}) on the die while keeping the clock rates in a range where
power consumption and heat dissipation do not represent a
problem. Multicore processors clearly represent the future of
computing. It is easy to imagine that multicore technologies will have
a deep impact on the High Performance Computing (HPC) world where high
processor counts are involved and, thus, limiting power consumption
and heat dissipation is a major requirement. The Top500~\cite{top500}
list released in June 2007 shows that the number of systems based on
the dual-core Intel Woodcrest processors grew in six months (i.e. from
the previous list) from 31 to 205 and that 90 more systems are based
on dual-core AMD Opteron processors.

Even if many attempts have been made in the past to develop
parallelizing compilers, they proved themselves efficient only on a
restricted class of problems. As a result, at this stage of the
multicore era, programmers cannot rely on compilers to take advantage
of the multiple execution units present on a processor. All the
applications that were not explicitly coded to be run on parallel
architectures must be rewritten with parallelism in mind. Also, those
applications that could exploit parallelism may need considerable
rework in order to take advantage of the fine-grain parallelism
features provided by multicores.

The current set of multicore chips from Intel and AMD are for the most
part multiple processors glued together on the same chip. There are
many scalability issues to this approach and it is unlikely that this
type of architecture will scale up beyond 8 or 16 cores. Even though
it is not yet clear how chip designers are going to address these
issues, it is possible to identify some properties that algorithms
must have in order to match high degrees of TLP:

\begin{description}
\item[fine granularity:] cores are (and probably will be) associated
  with relatively small local memories (either caches or explicitly
  managed memories like in the case of the STI
  Cell~\cite{isscc_2005_cell_desing} architecture or the Intel
  Polaris\cite{polaris} prototype). This requires splitting an
  operation into tasks that operate on small portions of data in order
  to reduce bus traffic and improve data locality. Moreover, for those
  architectures where cache memories are replaced by local memories,
  like the STI Cell processor, fine granularity is the only mean to
  achieve parallelism as suggested in previous work by the authors~\cite{cell_chol}.
\item[asynchronicity:] as the degree of TLP grows and the granularity of
  the operations becomes smaller, the presence of synchronization
  points in a parallel execution seriously affects the efficiency of
  an algorithm. Moreover, using asynchronous execution models it is
  possible to hide the latency of access to memory. The use of dynamic
  tasks execution was already studied in the past~\cite{76287,essl01}
\end{description}

Section~\ref{sec:lapack} shows why  such properties cannot be achieved
on algorithms  implemented in  commonly used linear  algebra libraries
due to their scalability limits in the context of multicore computing,
Section~\ref{sec:tiled}  describes fine granularity,  tiled algorithms
for the Cholesky, LU and  QR factorizations and presents a programming
model  for  their  asynchronous  and  dynamic  execution;  performance
results for this algorithm are shown in Section~\ref{sec:perf}.

%

\section{The LAPACK and ScaLAPACK libraries and their scalability
  limits}
\label{sec:lapack}

The LAPACK~\cite{lapack:99} and ScaLAPACK~\cite{scalapack:96} software
libraries represent a {\it de facto} standard for high performance
dense Linear Algebra computations and have been developed,
respectively, for shared-memory and distributed-memory
architectures\footnote{Here and in what follows, with LAPACK and
  ScaLAPACK we refer exclusively to the libraries reference
  implementations.}. In both cases exploitation of parallelism comes
from the availability of parallel BLAS.

The algorithms implemented in these two packages leverage the idea of
blocking to limit the amount of bus traffic in favor of a high reuse
of the data that is present in the higher level memories which are
also the fastest ones. This is achieved by recasting Linear Algebra
algorithms (like those implemented in LINPACK) in a way that the most
part of computations is done in Level-3 BLAS operations, where data
reuse is guaranteed by the so called {\it surface-to-volume} effect,
and only a small part in Level-2 BLAS for which memory bus speed
constitutes a performance upper bound~\cite{552704}.  As a result,
such algorithms can be roughly described as the repetition of two
fundamental steps:
\begin{description}
\item[panel factorization]: depending of the Linear Algebra operation
  that has to be performed, a number of transformations are computed
  for a small portion of the matrix (the so called {\it panel}). These
  transformations, computed by means of Level-2 BLAS operations, can
  be accumulated (the way they are accumulated changes depending on
  the particular operation performed).
\item[trailing submatrix update]: in this step, all the
  transformations that have been accumulated during the panel
  factorization, can be applied at once to the rest of the matrix
  (i.e. the trailing submatrix) by means of Level-3 BLAS operations.
\end{description}
Because the panel size is very small compared to the trailing
submatrix size, block algorithms are very rich in Level-3 BLAS
operations which provide high performance on memory hierarchy
systems. 

Both LAPACK and ScaLAPACK only exploit parallelism at the BLAS level,
i.e., by means of multithreaded BLAS libraries
(GotoBLAS~\cite{gotoblas}, MKL~\cite{mkl}, ATLAS~\cite{ATLAS},
ESSL\cite{essl01}, \dots) in the former case and by means of the
PBLAS~\cite{666023} library in the latter. Because Level-2 BLAS
operations cannot be efficiently parallelized on shared memory
(multicore) architectures due to the bus bottleneck, exploitation of
parallelism only at the BLAS level introduces a fork-join execution
pattern where:
\begin{itemize}
\item scalability is limited by the fact that the relative cost of
  strictly sequential operations (i.e., the panel factorization)
  increases when the degree of parallelism grows,
\item asynchronicity cannot be achieved because multiple threads are
  forced to wait in an idle state for the completion of sequential
  tasks. 
\end{itemize}

\begin{figure}[!h]
  \begin{center}
    \includegraphics[width=\textwidth]{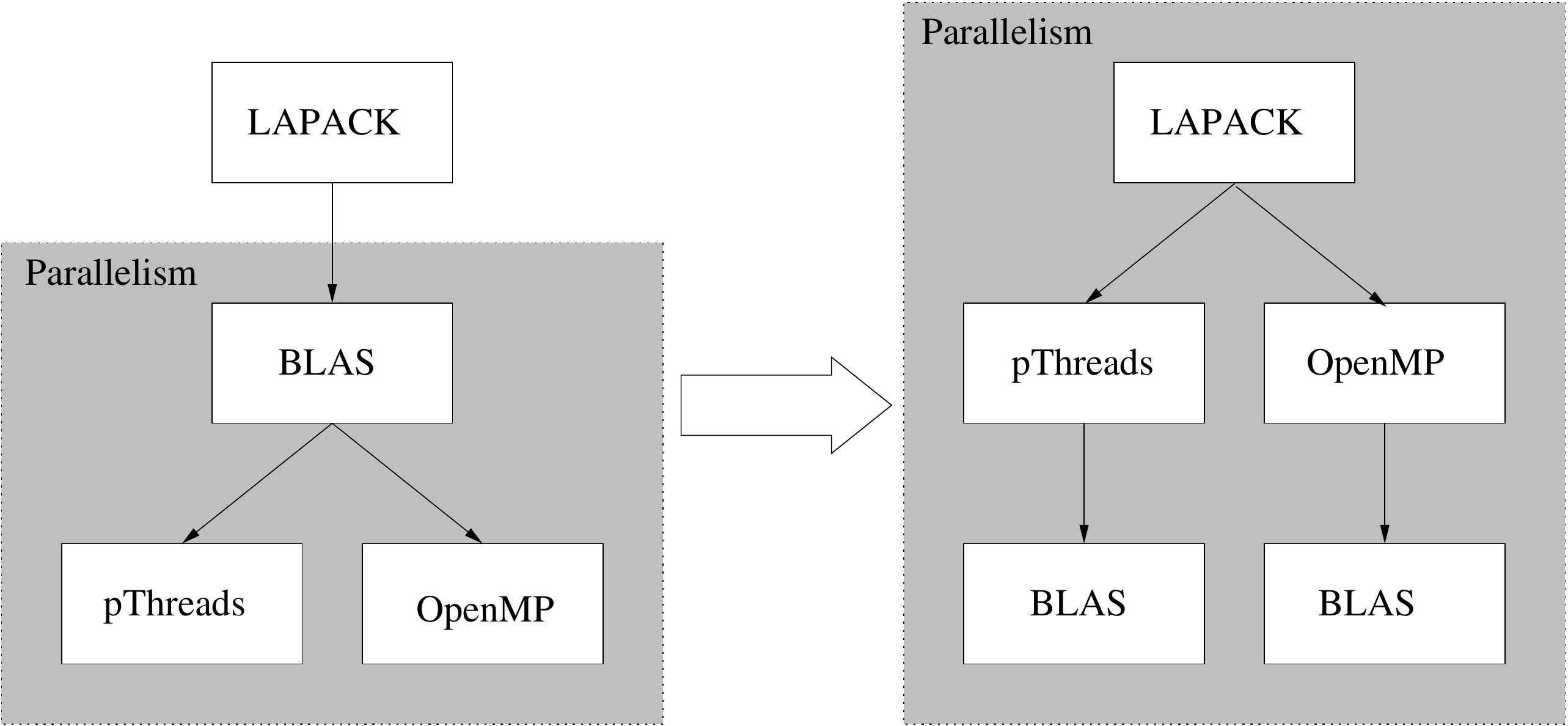}
  \caption{\label{fig:par_algos}Transition from sequential algorithms
    that rely on parallel BLAS to parallel algorithms.}
  \end{center}
\end{figure}

Algorithms for the QR, LU and Cholesky factorizations based on
recursion have been developed in the past~\cite{kag-gus,279535} in
order to increase the amount of computations performed in Level-3 BLAS
operations inside the panel. Even though they allow a better
exploitation of BLAS level parallelism, these algorithms are still not
suitable for achieving fine granularity levels.
 
As multicore systems require finer granularity and higher
asynchronicity, considerable advantages may be obtained by
reformulating old algorithms or developing new algorithms in a way
that their implementation can be easily mapped on these new
architectures by exploiting parallelism at an higher level. This
transition is shown in Figure~\ref{fig:par_algos}. Approaches along
these lines have already been studied in~\cite{76287,essl01} and, more
recently, by van de Geijn et al.~\cite{vdgqr,1248397} and the authors
of this paper~\cite{Kurzak:2006:ILA,para06,1248397, cell_chol}.

The technique described in~\cite{Kurzak:2006:ILA,para06} consists of
breaking the trailing submatrix update into smaller tasks that operate
on a block-column (i.e., a set of $b$ contiguous columns where $b$ is
the block size). The algorithm can then be represented as a Directed
Acyclic Graph (DAG) where nodes represent tasks, either panel
factorization or update of a block-column, and edges represent
dependencies among them. The execution of the algorithm is performed
by asynchronously scheduling the tasks in a way that dependencies are
not violated. This asynchronous scheduling results in an out-of-order
execution where slow, sequential tasks are hidden behind parallel
ones. 
Even if this approach provides significant speedup, as shown
in~\cite{Kurzak:2006:ILA, para06}, it is exposed to scalability
problems. Due to the relatively high granularity of the tasks, the
scheduling of tasks may have a limited flexibility and the parallel
execution of the algorithm may be affected by an unbalanced
load. Moreover, such a 1-D partitioning of the computational tasks is
not suited for such architectures like the Cell processor where memory
requirements impose a much smaller granularity.
The work described here aims at overcoming these limitations based on
the usage of the ``tiled'' algorithms described in
Section~\ref{sec:tiled}. 

The following sections describe the application of the idea of
dynamic scheduling and out of order execution to a class of algorithms
for Cholesky, LU and QR factorizations where finer granularity of the
operations and higher flexibility for the scheduling can be achieved. 
Fine granularity is obtained by using algorithms where the whole
factorization can be described as a sequence of tasks that operate on
small, square, portions of a matrix (i.e., tiles). Asynchronicity is
obtained by executing such algorithms according to a dynamic, graph
driven model.

\section{Fine Granularity Algorithms for the Cholesky, LU and QR Factorizations}
\label{sec:tiled}

As described in Section~\ref{sec:intro}, fine granularity is one of
the main requirements that is demanded to an algorithm in order to
achieve high efficiency on a parallel multicore system. This section
shows how it is possible to achieve this fine granularity for the
Cholesky, LU and QR factorizations by using ``tiled''
algorithms. Besides providing fine granularity, the use of tiled
algorithms also makes it possible to exploit more efficient storage
format for the data such as Block Data Layout (BDL). The benefits of
BDL have been extensively studied in the past, for example
in~\cite{670985,1014508}, and recent studies~\cite{1248397,tiledqr}
demonstrate how fine-granularity parallel algorithms can benefit from
BDL. A set of dense linear algebra algorithms for the BDL storage
format, was also introduced in the past by Gustavson et
al.~\cite{1014508,670985}.

Section~\ref{sec:async} shows how the idea of dynamic scheduling and
out of order execution, already discussed
in~\cite{Kurzak:2006:ILA,para06}, can be applied to these algorithms
in order to achieve the other important property described in
Section~\ref{sec:intro}, i.e. asynchronicity.  These ideas are not new
and have been proposed a number of times in the past~\cite{jordan,
  hep}.

\subsection{A Tiled Algorithm for the Cholesky Factorization}
Developing a tiled algorithm for the Cholesky factorization is a
relatively easy task since each of the elementary operations in the
standard LAPACK block algorithm can be broken into a sequence of
tasks that operate on small portions of data. The benefits of such
approach on parallel multicore systems have been already
discussed in the past~\cite{cell_chol,1248397,three,1014508}. 

The tiled algorithm for Cholesky factorization will be based on the
following set of kernel subroutines:
\begin{description}
\item[\texttt{DPOTF2}]. This LAPACK subroutine is used to perform the unblocked
  Cholesky factorization of a symmetric positive definite tile $A_{kk}$ of size $b \times
  b$ producing a unit, lower triangular tile $L_{kk}$. Thus, using the
  notation $input \longrightarrow output$, the call
  \texttt{DPOTF2}($A_{kk}$, $L_{kk}$) will perform
  \begin{displaymath}
    A_{kk} \longrightarrow L_{kk} = \textmd{Cholesky}(A_{kk})
  \end{displaymath}
\item[\texttt{DTRSM}]. This BLAS subroutine is used to apply the
  transformation computed by \texttt{DPOTF2} to a $A_{ik}$ tile by
  means of a triangular system solve. The \texttt{DTRSM}($L_{kk}$,
  $A_{ik}$, $L_{ik}$) performs
  \begin{displaymath}
    L_{kk}, A_{ik} \longrightarrow L_{ik} = A_{ik}L_{kk}^{-T}
  \end{displaymath}
\item[\texttt{DGSMM}]. This subroutine is used to update the tiles $A_{ij}$ in
  the trailing submatrix by mean of a matrix-matrix multiply. In the
  case of diagonal tiles, i.e. $A_{ij}$ tiled where $i=j$, this
  subroutine will take advantage of their triangular structure.
  The call \texttt{DGSMM}($L_{ik}$, $L_{jk}$, $A_{ij}$)
  \begin{displaymath}
    L_{ik}, L_{jk}, A_{ij} \longrightarrow A_{ij}=A_{ij}-L_{ik}L_{jk}^T
  \end{displaymath}
\end{description}

Assume a symmetric, positive definite matrix $A$ of size $n \times n$
where $n=p*b$ for some value $b$ that defines the size of the tiles
\begin{displaymath}
  A=\left(
    \begin{array}[!h]{cccc}
      A_{11} & 0     & \cdots & 0     \\
      A_{21} & A_{22} & \cdots & 0     \\
      \vdots& \vdots& \ddots & \vdots\\
      A_{p1} & A_{p2} & \cdots & A_{pp} \\
    \end{array}\right)
\end{displaymath}
where all the $A_{ij}$ are of size $b \times b$; then the tiled
Cholesky algorithm can be described as in Algorithm~\ref{alg:blkchol}.

\begin{algorithm}
\caption{\label{alg:blkchol}Tiled Cholesky factorization}
  \begin{algorithmic}[1]
    \FOR{k=1,...,p}
    \STATE \texttt{DPOTF2}($A_{kk}$, $L_{kk}$)
    \FOR{$i=k+1, ...,p$}
    \STATE \texttt{DTRSM}($L_{kk}$, $A_{ik}$, $L_{ik}$)
    \ENDFOR
    \FOR{$i=k+1, ...,p$}
    \FOR{$j=k+1, ...,i$}
    \STATE \texttt{DGSMM}($L_{ik}$, $L_{jk}$, $A_{ij}$)
    \ENDFOR
    \ENDFOR
    \ENDFOR
  \end{algorithmic}
\end{algorithm}

Note that no extra memory area is needed to store the $L_{ij}$ tiles
since they can overwrite the corresponding $A_{ij}$ tiles from the
original matrix.

\subsection{A Tiled Algorithm for the LU and QR Factorizations}

Although the same approach as for Cholesky can also be applied to the
LU and QR factorizations~\cite{670985}, this method, which consists of
simply rearranging the LAPACK algorithm in terms of operations by
tiles, incurs into efficiency problems due to the fact that, in a
panel factorization step, each of the tiles that compose the panel is
accessed multiple times. 

For this reason we propose an algorithmic change which takes its roots
in updating factorizations~\cite{golubvanloan,stew:98}.  Using
updating techniques to tile the algorithms have first\footnote{to our
  knowledge} been proposed by Yip ~\cite{yip_ooc} for LU to improve
the efficiency of out-of-core solvers, and were recently reintroduced
in~\cite{DBLP:conf/para/JoffrainQG04,vdgooclu,1055534} for LU and QR, once more in the out-of-core
context. A similar idea has also been proposed in~\cite{210517} for
Hessenberg reduction in the parallel distributed context. The
efficiency of these algorithms in a parallel multicore system has been
discussed, for the QR factorization, in~\cite{tiledqr}; specifically
the algorithm used in~\cite{tiledqr} is a simplified variant of that
discussed in~\cite{1055534} that aims at overcoming the limitations of
BLAS libraries on small size tiles. The cost of this simplification is
an increase in the operation count for the whole QR factorization. In
this document the same algorithm as in~\cite{1055534} is used to
achieve high efficiency for both the LU and QR factorizations;
performance results show that this choice, while limiting the
operation count overhead to a negligible amount, still delivers high
execution rates. This approach has been presented for the QR
factorization in~\cite{vdgqr}.

A stability analysis for the tiled  algorithm for LU factorization may be found
in~\cite{vdgooclu}.

\subsubsection{Tiled Algorithm for the QR Factorization}
The description of the tiled algorithm for the QR factorization will be based on
the following sets of kernel subroutines:

\begin{description}
\item[\texttt{DGEQRT}.] This subroutine was developed to perform the
  block QR factorization of a diagonal
  block $A_{kk}$ of size $b \times b$ with internal block size
  $s$. This operation produces an upper triangular matrix $R_{kk}$, a
  unit lower triangular matrix $V_{kk}$ that contains $b$ Householder
  reflectors and an upper triangular matrix $T_{kk}$ as defined by the
  compact WY technique for accumulating Householder
  transformations~\cite{64889}. This kernel subroutine is based on the
  LAPACK \texttt{DGEQRF} one and, thus, it consists mostly of Level-3
  BLAS operations; in addition to the LAPACK subroutine,
  \texttt{DGEQRT} also computes the $T_{kk}$ matrix.

  Thus, using the notation $input \longrightarrow output$, the call
  \texttt{DGEQRT}($A_{kk}$, $V_{kk}$, $R_{kk}$, $T_{kk}$) performs 
  \begin{displaymath}
    A_{kk} \longrightarrow (V_{kk}, R_{kk}, T_{kk})=QR(A_{kk})
  \end{displaymath}

\item[\texttt{DLARFB}.] This LAPACK subroutine, based exclusively on
  Level-3 BLAS operations, will be used to apply
  the transformation $(V_{kk}, T_{kk})$ computed by subroutine
  \texttt{DGEQRT} to a tile $A_{kj}$ producing a $R_{kj}$ tile.

  Thus, \texttt{DLARFB}($A_{kj}$, $V_{kk}$, $T_{kk}$, $R_{kj}$) performs
  \begin{displaymath}
    A_{kj}, V_{kk}, T_{kk} \longrightarrow R_{kj}=(I-V_{kk}T_{kk}V^T_{kk})A_{kj}
  \end{displaymath}

\item[\texttt{DTSQRT}.] This subroutine was developed to perform the
  blocked QR factorization of a matrix that is formed by coupling an
  upper triangular block $R_{kk}$ with a square block $A_{ik}$ with
  internal block size $s$. This subroutine will return an upper
  triangular matrix $R_{kk}$, an upper triangular matrix
  $T_{ik}$ as defined by the compact WY technique for accumulating householder
  transformations, and a tile $V_{ik}$ containing $b$ Householder
  reflectors where $b$ is the tile size.

  Then, \texttt{DTSQRT}($R_{kk}$, $A_{ik}$, $V_{ik}$, $T_{ik}$) performs
  \begin{displaymath}
    \left(
      \begin{array}[!h]{c}
        R_{kk}\\
        A_{ik}
      \end{array}\right) \longrightarrow
        (V_{ik}, T_{ik}, R_{kk})=QR\left(\begin{array}[!h]{c}
        R_{kk}\\
        A_{ik}
      \end{array}\right)
  \end{displaymath}

\item[\texttt{DSSRFB}.] This subroutine was developed to update the
  matrix formed by coupling two square blocks $R_{kj}$ and $A_{ij}$ applying the
  transformation computed by \texttt{DTSQRT}. 

  Thus, \texttt{DSSRFB}($R_{kj}$, $A_{ij}$, $V_{ik}$, $T_{ik}$)
  performs
  \begin{displaymath}
    \left(\begin{array}{c}
      R_{kj}\\
      A_{ij}
    \end{array}\right), V_{ik}, T_{ik} \longrightarrow
    \left(
    \begin{array}{c}
      R_{kj}\\
      A_{ij}\\
    \end{array}\right)=
    (I - V_{ik}T_{ik} V^T_{ik})\left(\begin{array}{c}
      R_{kj}\\
      A_{ij}
    \end{array}\right)
  \end{displaymath}
\end{description}

Note that no extra storage is required for the $V_{ij}$ and $R_{ij}$
since those tiles can overwrite the $A_{ij}$ tiles of the original
matrix $A$; a temporary memory area has to be allocated to store the
$T_{ij}$ tiles. Further details on the implementation of the \texttt{DTSQRT}
and \texttt{DSSRFB} are provided in Section~\ref{sec:cost}.

Assuming a matrix $A$ of size $pb \times qb$
\begin{displaymath}
  \left(\begin{array}[!h]{cccc}
    A_{11}  & A_{12} & \dots  & A_{1q}  \\
    A_{21}  & A_{22} & \dots  & A_{2q}  \\
    \vdots &       & \ddots & \vdots \\
    A_{p1}  & A_{p2} & \dots  & A_{pq}
  \end{array}\right)
\end{displaymath}
where $b$ is the block size and each $A_{ij}$ is of size $b \times b$,
the QR factorization can be performed as in Algorithm~\ref{alg:blkqr}.

\begin{algorithm}[!h]
\caption{\label{alg:blkqr}The tiled algorithm for QR factorization.}
  \begin{algorithmic}[1]
    \FOR{$k=1, ..., min(p,q)$}
    \STATE \texttt{DGEQRT}($A_{kk}$, $V_{kk}$, $R_{kk}$, $T_{kk}$)
    \FOR{$j=k+1, ..., q$}
    \STATE \texttt{DLARFB}($A_{kj}$, $V_{kk}$, $T_{kk}$, $R_{kj}$)
    \ENDFOR
    \FOR{$i=k+1, ..., p$}
    \STATE \texttt{DTSQRT}($R_{kk}$, $A_{ik}$, $V_{ik}$, $T_{ik}$)
    \FOR{$j=k+1, ..., q$}
    \STATE \texttt{DSSRFB}($R_{kj}$, $A_{ij}$, $V_{ik}$, $T_{ik}$)
    \ENDFOR
    \ENDFOR
    \ENDFOR
  \end{algorithmic}
\end{algorithm}

\subsubsection{Tiled Algorithm for the LU Factorization}

The description of the tiled algorithm for the LU factorization will be based on
the following sets of kernel subroutines.

\begin{description}
\item[\texttt{DGETRF}.] This LAPACK subroutine, consisting mostly of
  Level-3 BLAS operations, performs a block LU
  factorization of a tile $A_{kk}$ of size $b \times b$ with internal block
  size $s$. As a result, two matrices $L_{kk}$ and $U_{kk}$, unit-lower and
  upper triangular respectively, and a permutation matrix $P_{kk}$ are
  produced. Thus, using the notation $input \longrightarrow output$,
  the call \texttt{DGETRF}($A_{kk}$, $L_{kk}$, $U_{kk}$, $P_{kk}$) will perform 
  \begin{displaymath}
    A_{kk} \longrightarrow L_{kk},U_{kk}, P_{kk} = LU(A_{kk})
  \end{displaymath}
\item[\texttt{DGESSM}.] This routine, based on Level-3 BLAS
  operations, was developed to apply the transformation
  $(L_{kk},P_{kk})$ computed by the \texttt{DGETRF} subroutine to a
  tile $A_{kj}$. thus the call \texttt{DGESSM}($A_{kj}$, $L_{kk}$,
  $P_{kk}$, $U_{kj}$) will perform
  \begin{displaymath}
    A_{kj}, L_{kk}, P_{kk} \longrightarrow U_{kj}=L_{kk}^{-1}P_{kk}A_{kj}
  \end{displaymath}
\item[\texttt{DTSTRF}.] This subroutine was developed to perform the
  block LU factorization of a matrix that is formed by coupling the
  upper triangular block $U_{kk}$ with a square block $A_{ik}$ with
  internal block size $s$. This subroutine will return an upper
  triangular matrix $U_{kk}$,
  a unit, lower triangular matrix $L_{ik}$ and a permutation matrix
  $P_{ik}$. Thus, the call \texttt{DTSTRF}($U_{kk}$, $A_{ik}$, $P_{ik}$) will perform
  \begin{displaymath}
    \left(
      \begin{array}{c}
        U_{kk}\\
        A_{ik}
      \end{array}\right) \longrightarrow U_{kk}, L_{ik}, P_{ik}
    = LU \left(
      \begin{array}{c}
        U_{kk}\\
        A_{ik}
      \end{array}\right)
  \end{displaymath}

\item[\texttt{DSSSSM}.] This subroutine was developed to update the
  matrix formed by coupling two square blocks $U_{kj}$ and $A_{ij}$ applying the
  transformation computed by \texttt{DTSTRF}. Thus the call
  \texttt{DSSSSM}($U_{kj}$, $A_{ij}$, $L_{ik}$, $P_{ik}$) performs
  \begin{displaymath}
    \left(
      \begin{array}{c}
        U_{kj}\\
        A_{ij}
      \end{array}\right), L_{ik}, P_{ik} \longrightarrow \left(\begin{array}{c}
        U_{kj}\\
        A_{ij}
      \end{array}\right)= L_{ik}^{-1}P_{ik} \left(\begin{array}{c}
        U_{kj}\\
        A_{ij}
      \end{array}\right)
  \end{displaymath}
\end{description}

Note that no extra storage is required for the $U_{ij}$ since they can
overwrite the correspondent $A_{ij}$ tiles of the original matrix
$A$. A memory area must be allocated to store the $P_{ij}$ and part of
the $L_{ij}$; the $L_{ij}$ tiles, in fact, are $2b \times b$ matrices,
i.e. two tiles arranged vertically and, thus, one tile can overwrite
the corresponding $A_{ij}$ tile and the other is stored in the extra
storage area\footnote{the upper part of $L_{ij}$ is, actually, a group
  of $b/s$ unit, lower triangular matrices each of size $s \times s$
  and, thus, only a small memory area is required to store it.}.
Further details on the implementation of the \texttt{DTSTRF} and
\texttt{DSSSSM} are provided in Section~\ref{sec:cost}.

Assuming a matrix $A$ of size $pb \times qb$
\begin{displaymath}
  \left(\begin{array}[!h]{cccc}
    A_{11}  & A_{12} & \dots  & A_{1q}  \\
    A_{21}  & A_{22} & \dots  & A_{2q}  \\
    \vdots &       & \ddots & \vdots \\
    A_{p1}  & A_{p2} & \dots  & A_{pq}
  \end{array}\right)
\end{displaymath}
where $b$ is the block size and each $A_{ij}$ is of size $b \times b$,
the LU factorization can be performed as in Algorithm~\ref{alg:blklu}.

\begin{algorithm}[!h]
\caption{\label{alg:blklu}The tiled algorithm for LU factorization.}
  \begin{algorithmic}[1]
    \FOR{$k=1, ..., min(p,q)$}
    \STATE \texttt{DGETRF}($A_{kk}$, $L_{kk}$, $U_{kk}$, $P_{kk}$)
    \FOR{$j=k+1, ..., q$}
    \STATE \texttt{DGESSM}($A_{kj}$, $L_{kk}$, $P_{kk}$, $U_{kj}$)
    \ENDFOR
    \FOR{$i=k+1, ..., p$}
    \STATE \texttt{DTSTRF}($U_{kk}$, $A_{ik}$, $P_{ik}$)
    \FOR{$j=k+1, ..., q$}
    \STATE \texttt{DSSSSM}($U_{kj}$, $A_{ij}$, $L_{ik}$, $P_{ik}$)
    \ENDFOR
    \ENDFOR
    \ENDFOR
  \end{algorithmic}
\end{algorithm}

\begin{figure}[!h]
  \begin{center}
    \includegraphics[width=\textwidth]{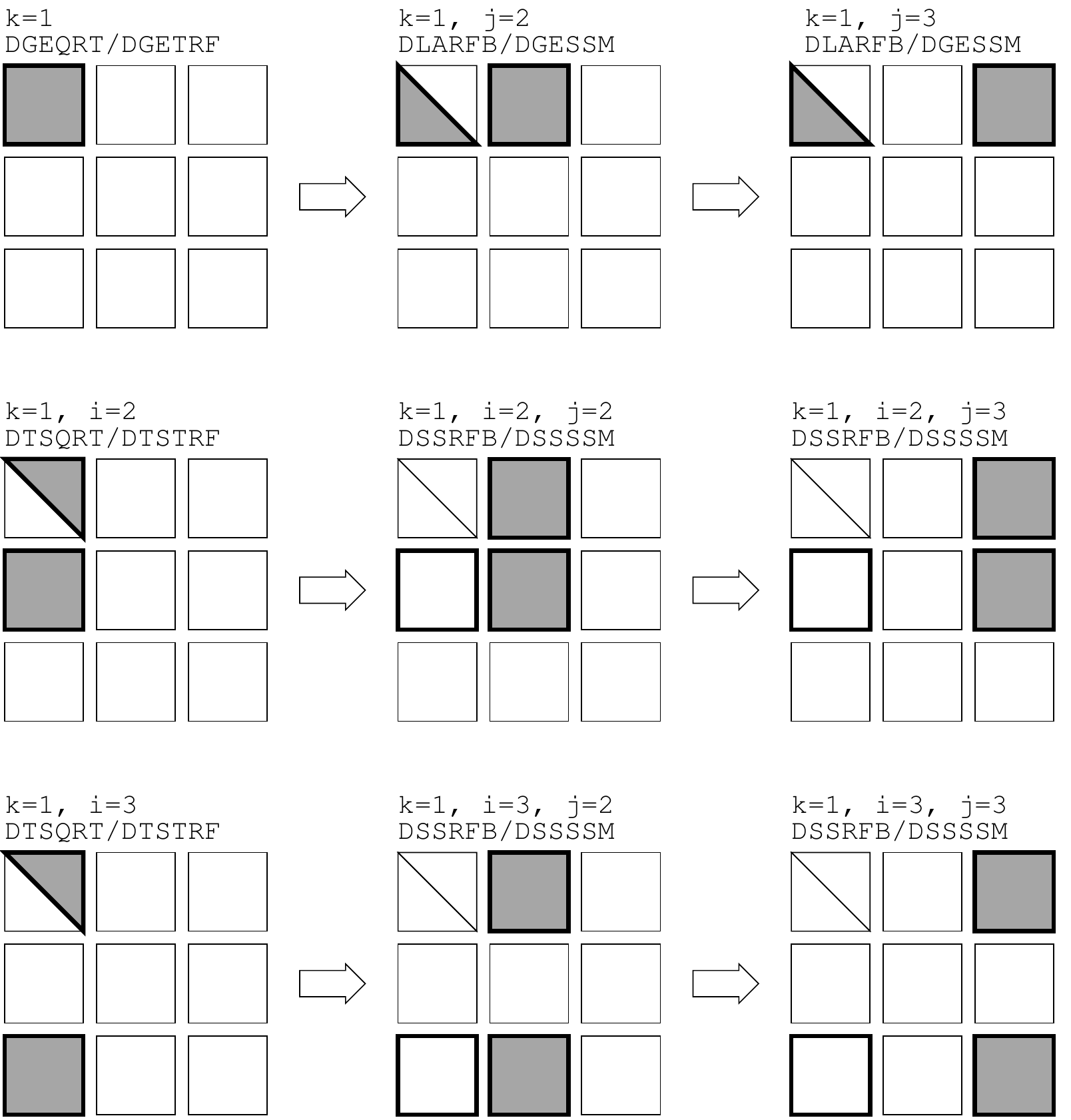}
    \caption{\label{fig:blk_alg}Graphical representation of one
      repetition of the outer loop in Algorithms~\ref{alg:blkqr}
      and~\ref{alg:blklu} on a matrix with $p=q=3$. A thick border
      shows the tiles that are being read and a gray fill shows the
      tiles that are being written at each step.  As expected the
      picture is very similar to the out-of-core algorithm presented
      in~\cite{1055534}.  }
\end{center}
\end{figure}

Since the only difference between Algorithms~\ref{alg:blkqr}
and~\ref{alg:blklu} is in the kernel subroutines, and noting, as
explained before, that the $R_{ij}$, $V_{ij}$, $U_{ij}$ and $L_{ij}$
tiles are stored in the corresponding memory locations that contain
the tiles $A_{ij}$ of the original matrix $A$ (the $L_{ij}$ only
partially), a graphical representation of Algorithms~\ref{alg:blkqr}
and~\ref{alg:blklu} is as in Figure~\ref{fig:blk_alg}.

\subsubsection{Reducing the Cost of the Tiled Algorithms for the QR
  and LU Factorization}
\label{sec:cost}

Because the gap between processor and memory speeds is likely to
increase with multicore technologies, the usage of blocking
transformations is of great importance to achieve high data reuse in
linear algebra operations. However, blocking of transformations
introduces an extra cost in the operation count of the tiled
algorithms for the QR and LU
factorizations~\cite{DBLP:conf/para/JoffrainQG04,vdgooclu,1055534,tiledqr,vdgqr}. In this section
we describe a method, presented in~\cite{DBLP:conf/para/JoffrainQG04,vdgooclu,1055534,vdgqr}, to
keep this extra cost limited to a negligible amount. Since this method
applies identically to the tiled algorithms for both the QR and LU
factorizations, only the former case is treated in the following
discussion.

Based on the observation that the \texttt{DGEQRT}, \texttt{DLARFB} and
\texttt{DTSQRT} kernels only contribute lower order terms (only
$O(n^2)$, $n$ being the size of the problem), the cost of
the tiled algorithm for the QR factorization is determined by the cost
of the \texttt{DSSRFB} kernel. It is, thus, important to pay attention
to the way the transformations applied by \texttt{DSSRFB} are computed
and accumulated in \texttt{DTSQRT}. The method presented
in~\cite{DBLP:conf/para/JoffrainQG04,vdgooclu,1055534,tiledqr,vdgqr} suggests that these
transformations can be accumulated in sets of $s$; assuming $s \ll b$,
where $b$ is the tile size, the extra cost introduced by blocking is
limited to a negligible amount, as demonstrated below.
\begin{figure}[!h]
  \begin{center}
    \includegraphics[width=\textwidth]{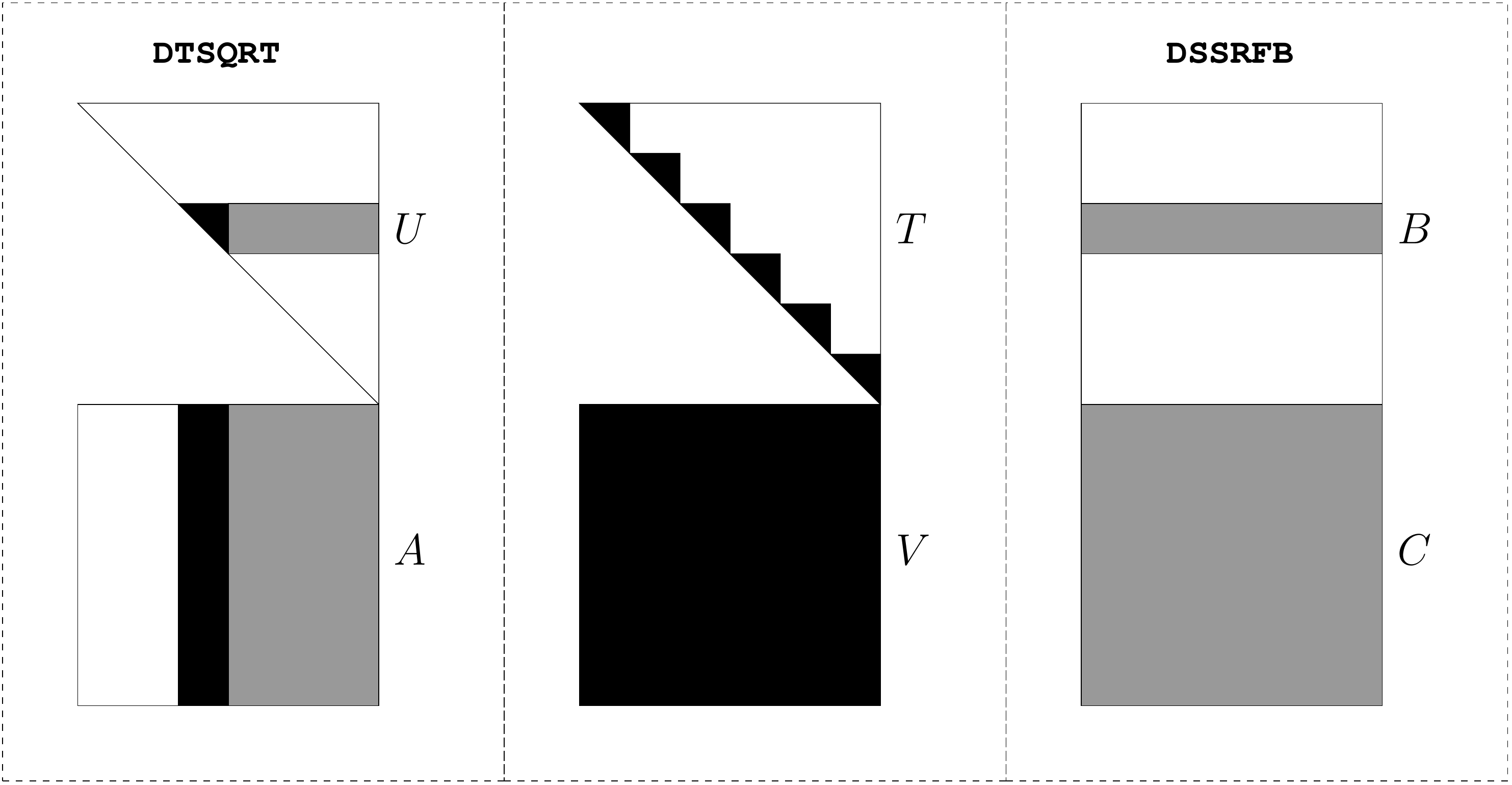}
  \caption{\label{fig:inner_blk}Details of the computation and
    accumulation of transformations in \texttt{DTSQRT} and their
    application in \texttt{DSSRFB}.}
  \end{center}
\end{figure}
This technique is illustrated in Figure~\ref{fig:inner_blk}. Assuming
$b/s = t$, the \texttt{DTSQRT}($U$, $A$, $V$, $T$) performs a loop of
$t$ repetitions where, at each step $i$, a set of $s$ Householder
reflectors $V_{i}=(v_{i1} v_{i2} \dots v_{is})$ are computed and
accumulated according to the $WY$ technique already mentioned
above. As a result of the accumulation, an upper triangular matrix
$T_i$ of size $s\times s$ is formed ( $V_i$ and $T_i$ are highlighted
in black in Figure~\ref{fig:inner_blk}({\it center})). This amounts to
a blocking QR factorization (with block size $s$) of the couple formed
by the $U$ and $A$ tiles (in Figure~\ref{fig:inner_blk} ({\it left})
the panel and the trailing submatrix are highlighted in black and
grey, respectively).  By the same token, the \texttt{DSSRFB}($B$, $C$,
$V$, $T$) performs a loop of $t$ repetitions where, at step $i$, a
portion of the $B$ and $C$ tiles is updated by the application of the
transformations computed in \texttt{DTSQRT} and accumulated in $V_i$
and $T_i$. The data updated at each step of \texttt{DSSRFB} is
highlighted in grey in Figure~\ref{fig:inner_blk} ({\it right}).

The cost for a single call of the \texttt{DSSRFB} kernel is, ignoring
the lower order terms, $4b^3 + sb^2$; consequently, the cost of
the whole QR factorization is
\begin{displaymath}
  \sum_{k=1}^q (4b^2+sb^2)(p-k)(q-k) \simeq 2n^2(m-\frac{n}{3})(1+\frac{s}{4b})
\end{displaymath}
assuming that $q<p$ and that $p$ and $q$ are big enough so that it is
possible to ignore the $O(n^2)$ contributions from the
\texttt{DGEQRT}, \texttt{DLARFB} and \texttt{DTSQRT} kernels. It must
be noted that, when $s=b$, the cost of the tiled algorithm is 25\%
higher than that of the standard LAPACK one; the choice $s=b$ may help
overcoming the limitations of commonly used BLAS libraries on small
size data~\cite{tiledqr} but, as performance results show (see
Section~\ref{sec:perf}) it is possible to define values for $b$ and
$s$ capable of reducing the extra cost to a negligible amount while
providing a good level of performance.

This tile level blocking technique can be applied to the
\texttt{DTSRFT} and \texttt{DSSSSM} kernel subroutines for the tiled
LU factorization as well. This leads to a cost of $2b^3+sb^2$ for the
\texttt{DSSSSM} kernel and
\begin{displaymath}
  \sum_{k=1}^q (2b^2+sb^2)(p-k)(q-k) \simeq n^2(m-\frac{n}{3})(1+\frac{s}{2b})
\end{displaymath}
for the whole factorization under the same assumption as before.

It has to be noted that, too small values for $s$ may hurt the
performance of the Level-3 BLAS operations used in the kernel
subroutines. It is, thus, very important to carefully choose the
correct values for $b$ and $s$ that offer the better compromise
between extra cost minimization and efficiency of Level-3 BLAS
operations.

\subsubsection{Stability of the Tiled Algorithm for the LU Factorization}
\label{sec:stability}

Algorithm~\ref{alg:blklu} performs eliminations with different pivots than \textit{Gaussian elimination with partial pivoting} (GEPP).
For eliminating the $(n-k)$ entries in column $k$, partial pivoting chooses a unique pivot while,
Algorithm~\ref{alg:blklu} potentially uses up to $(n-k)/b$ pivots. 
The pivoting strategy considered in Algorithm~\ref{alg:blklu} is indeed a tiled version of
\textit{Gaussian elimination with pairwise pivoting} (GEWP)
where GEPP is used at the block level. For this reason,
we call the pivoting strategy used by Algorithm~\ref{alg:blklu}: \textit{Gaussian elimination with tiled pairwise pivoting} (GETWP).
When $b=1$ (a $n$--by--$n$ tiled matrix with $1$--by--$1$ tiles), GETWP reduces to GEWP.
When $b=n$ (a $1$--by--$1$ tiled matrix with $n$--by--$n$ tiles), GETWP reduces to GEPP.

GEWP dates back to Wilkinson's work~\cite{wilk:65}.
Wilkinson's motivation was to cope with limited amount of memory in contemporary computers.
The
approach has been since successfully used in 
out-of-core solvers (e.g.~\cite{DBLP:conf/para/JoffrainQG04,vdgooclu,reid:71,yip_ooc}) or in
the parallel context (see \cite[\S4.2.2]{gaps:90} for a summary of references).

The stability analysis of GEPP is not
well understood, the accepted idea is that GEPP is \textit{practically stable}. It is only our experience that
makes us conjecture that the practical behavior is stable and indeed far different from
a few contrived unstable examples~\cite{hihi:89,trsc:90}.

Unfortunately and unsurprisingly, the stability analysis of GETWP
is as badly understood. In this
section, we build experience with this pivoting strategy and conclude that
\begin{enumerate}
\item GETWP is less stable than GEPP; the smaller $b$ is, the less stable the method is;
\item we highly recommend to check the backward error for the solution after
a linear solve (i.e. do not trust the answer, check it) and perform a
few steps of iterative refinement if needed;
\item our observations have lead to more questions than answers.
\end{enumerate}

GEPP and GETWP consists in the successive applications onto $A$ of both an elementary unit lower triangular matrix ($L$)
and an elementary permutation matrix ($P$).
For GEPP, there are $(n-1)$ couples and we write
\begin{equation}\label{eq:GEPP} U  = L_{n-1} P_{n-1} \ldots  L_1 P_1 A.\end{equation}
For GETWP, with $p=n/b$, there are $(p)(p-1)/2$ couples and we write:
\begin{equation}\label{eq:GETWP} 
U  = 
  L_{p,p} P_{p,p}  
  L_{p,p-1} P_{p,p-1} L_{p-1,p-1} P_{p-1,p-1}
  \ldots
  L_{2,p} P_{2,p} \ldots  L_{2,2} P_{2,2} 
  L_{1,p} P_{1,p} \ldots  L_{1,1} P_{1,1} A.
\end{equation}

In GEPP and GETWP, a pivot can eliminate an element only if
the eliminator (pivot) is larger in absolute value than the eliminatee. 
Consequently, the multipliers
(the off-diagonal elements of $L_k$ (GEPP) or $L_{i,j}$ (GETWP)) are
smaller or equal than $1$ in absolute value.

In the case of GEPP, Equation~(\ref{eq:GEPP}) can be rearranged in the form:
\begin{equation}\label{eq:GEPPcompact} 
 L U = P A, 
\end{equation}
where $P = P_{n-1}\ldots P_1$ and $L$ is obtained
by taking nonzeros off-diagonal elements in the elementary transformations $L_k$,
changing their signs, and applying the permutations accordingly.
Therefore, in GEPP, all the entries below the diagonal of $L$ are
smaller than $1$ in absolute value. Consequently, the norm of $L$ is bounded independently of $A$,
we have $\| L \|_\infty \leq n$. This observation is crucial in the study of the stability 
of GEPP and explains the focus in the literature on $\| U \|_\infty/\|A\|_\infty$.

In the case of GETWP, we define
\begin{equation}\label{eq:formula N}
N = 
\left(
  L_{p,p} P_{p,p}  
  L_{p,p-1} P_{p,p-1} L_{p-1,p-1} P_{p-1,p-1}
  \ldots
  L_{2,p} P_{2,p} \ldots  L_{2,2} P_{2,2} 
  L_{1,p} P_{1,p} \ldots  L_{1,1} P_{1,1}
\right)^{-1},
\end{equation}
so that we can write from Equation~(\ref{eq:GETWP}): 
\begin{equation}\label{eq:GETWPcompact} 
N U = A. 
\end{equation}
Equation~(\ref{eq:GETWPcompact}) is to GETWP what 
Equation~(\ref{eq:GEPPcompact}) is to GEPP.
We note that $N$ is a mathematical artifact and in practice $N$ is not computed but manipulated through $L_{i,k}$ and $P_{i,k}$.

Three main differences occur between $L$ from GEPP and $N$ from GETWP:
\begin{enumerate}
\item $N$ is the combination of permutations and elementary transformations, the effect of which can not be dissociated
 in two matrices $L$ and $P$ as it is for GEPP , this complicates notably any analysis,
\item although we need $n(n-1)/2$ elements to store all the $\left(L_{i,j}\right)$'s, the matrix $N$ is not unit lower triangular
and has in general a lot more than $n(n-1)/2$ entries,
\item the absolute values of the entries of the off-diagonal elements of $N$
can be greater than $1$ and in practice they are notably larger, therefore a
stability analysis of GETWP requires us not only to monitor $\|U\|_\infty$ but
also $\|L\|_\infty$.
\end{enumerate}

In term of related theoretical work, Sorensen~\cite{sore:85} has proved that the
worst case behavior for the growth in $U$ for GETWP is $2^{n-1}$ (same bound for GEPP)
while the worst case behavior for the growth in $L$ is $2^{n-1}$ (GEPP is
$\sqrt{n}$). We know that these worth case scenarios come from contrived examples
and so our present analysis tries to clarify what the general case behavior is.

Experimental results of the stability of GEWP are given in~\cite{trsc:90} where
Trefethen and Schreiber experimentally showed that the growth factor in $U$ is
smaller than $n$ for a set of random matrices ($n \leq 1024$).
Quintana-Ort\'{\i} and van de Geijn~\cite{vdgooclu} have experimentally studied
GETWP on random matrices with two tiles (case $b=n/2$).

The present section details results for GETWP with various block sizes on
matrices coming from random matrices and applications.

\paragraph{Random matrices.}

We take 10 random matrices of size $n=2048$ (\texttt{A=randn(n)}). Any reported
quantities reported is indeed the mean obtained from this sample.

To evaluate the backward error for the factorization of GETWP, we need to compute the  $N$ factor.
From Equation (\ref{eq:formula N}), we get
$$
N = 
 P_{1,1}^{-1}
 L_{1,1}^{-1}
 \ldots
 P_{1,p}^{-1}
 L_{1,p}^{-1}
 P_{2,2} ^{-1}
 L_{2,2}^{-1}
 \ldots 
 P_{2,p}^{-1}
 L_{2,p}^{-1}
 \ldots
 P_{p-1,p-1}^{-1}
 L_{p-1,p-1}^{-1}
 P_{p,p-1}^{-1}
 L_{p,p-1}^{-1}
 P_{p,p}  ^{-1}
 L_{p,p}^{-1}.
$$

\begin{figure}
\begin{minipage}{0.45 \textwidth}
\includegraphics[width=1.00 \textwidth]{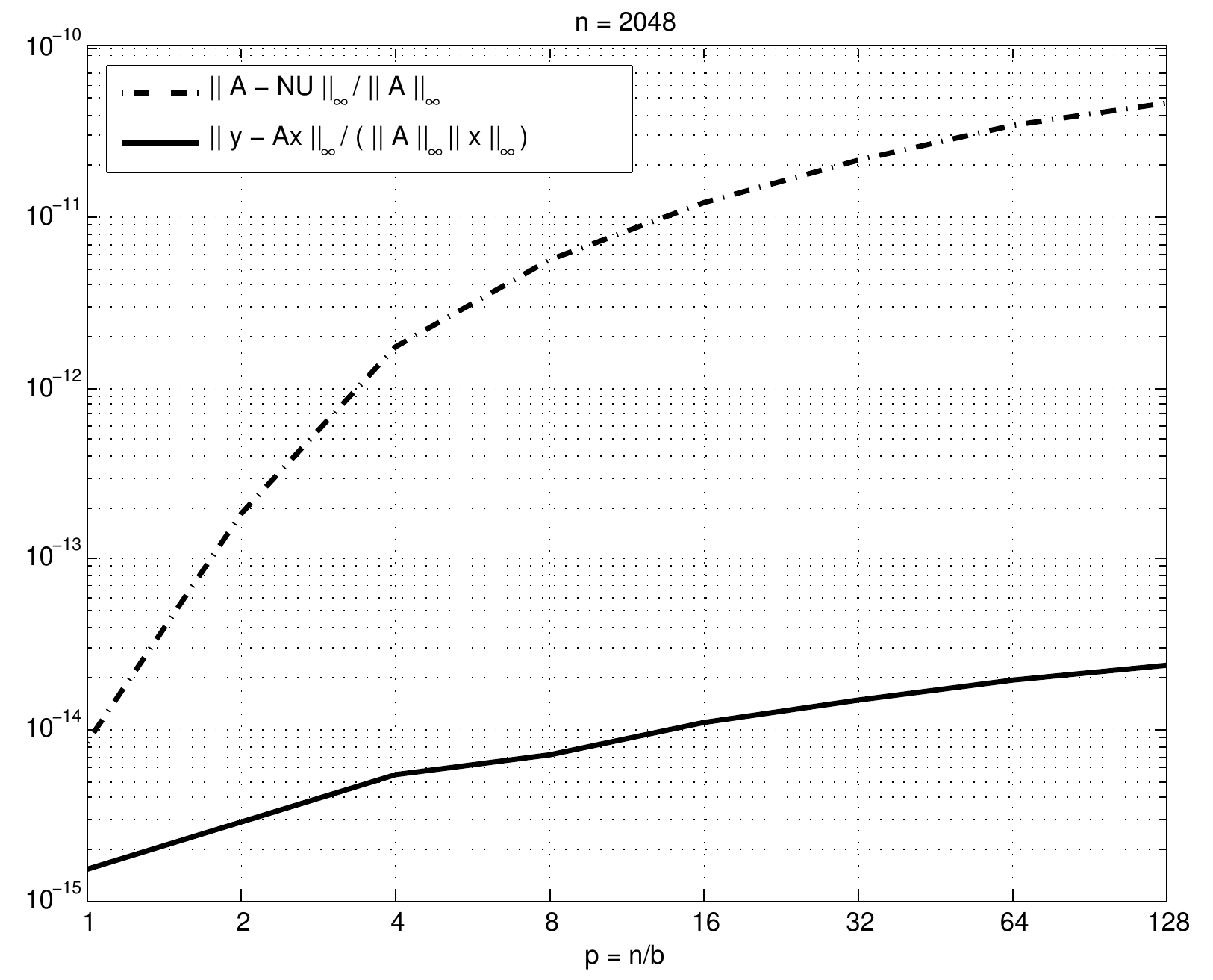}
\end{minipage}
\begin{minipage}{0.45 \textwidth}
\includegraphics[width=1.00 \textwidth]{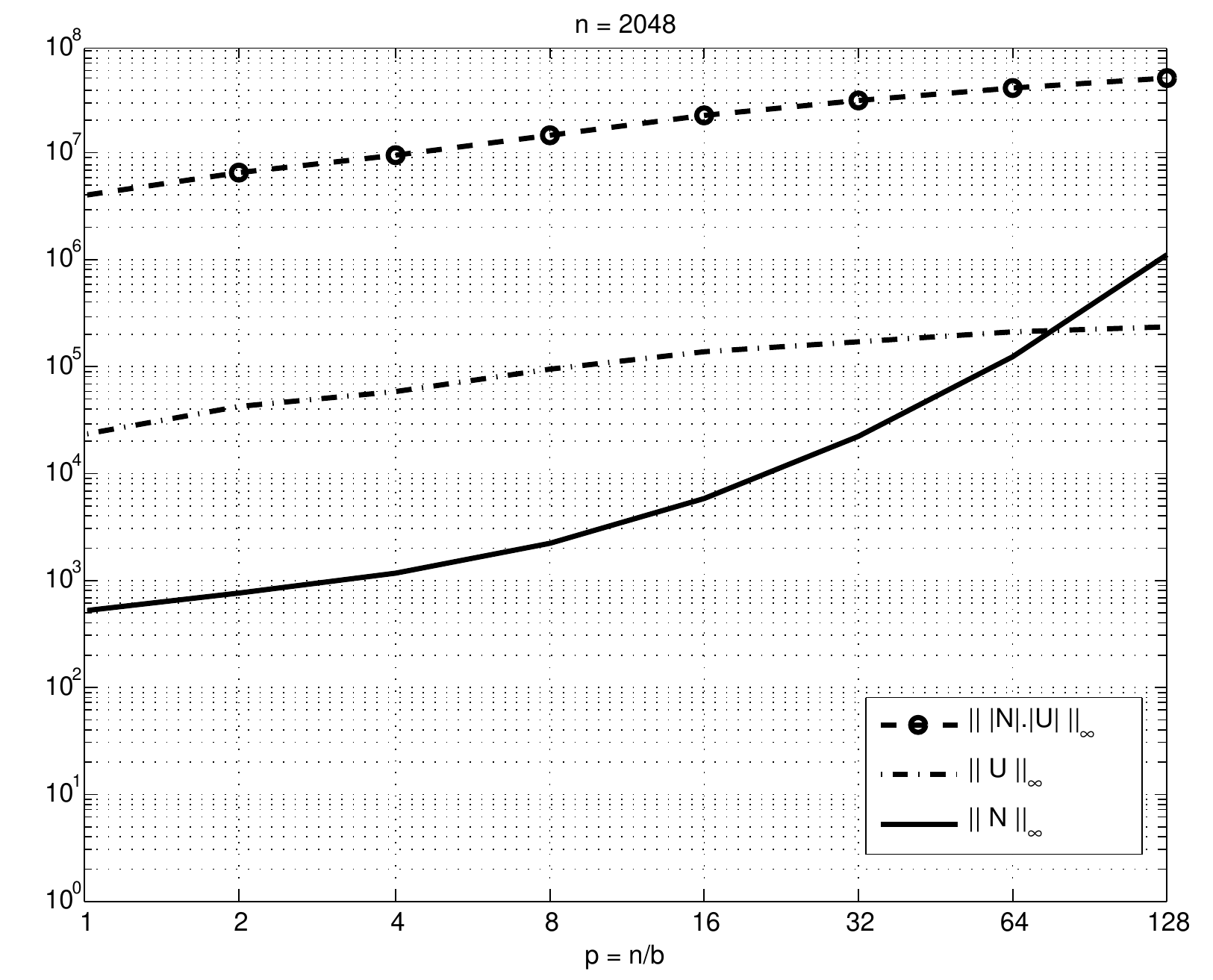}
\end{minipage}
\caption{\label{fig:random_nb_n2048}
We take a sample of 10 random matrices of size $n=2048$.
On the left, we plot the mean backward error obtained for the GETPW factorization and
the mean backward error for the solution when solving a linear system of equations with the GETPW factorization
with a random right-hand side. In the x-axis, we make the tile size ($b$) decrease
from $b=n=2048$ (corresponds to $p=1$) to $b=16$ (corresponds to $p=128$).
On the right, we plot the mean of various relevant quantities for the GETPW factorization.
}
\end{figure}

On the left of Figure~\ref{fig:random_nb_n2048}, we plot the backward error for the factorization obtained with GETPW
$$ \left(\frac{ \| A - N_{\textsc{\tiny wp}}U_{\textsc{\tiny wp}} \|_\infty}{\| A \|_\infty }\right) $$
and
the backward error for the solution when solving a linear system of equations with the GETPW factorization and
a random right-hand side
$$ \left(\frac{ \| y - A x_{\textsc{\tiny wp}} \|_\infty}{\| A \|_\infty \|x_{\textsc{\tiny wp}}\|_\infty }\right). $$
The horizontal axis represents various numbers of tiles ($p$).
For $p=1$, there is one tile so the algorithm is indeed GEPP.
For $p=2$, there are four 1024--by--1024 tiles, etc.
As the number of tiles increases, the stability of GETWP decreases.
We note that there is a significant difference between the backward error for the solution
and the backward error for the factorization.

On the right of Figure~\ref{fig:random_nb_n2048}, we plot the three quantities:
$$
 \| N_{\textsc{\tiny wp}} \|_\infty,\quad
 \| U_{\textsc{\tiny wp}} \|_\infty,\quad \textmd{and}\quad
 \| |N_{\textsc{\tiny wp}} |\cdot| U_{\textsc{\tiny wp}} | \|_\infty.
$$
The relevant quantity for the stability of the factorization being $\| |N_{\textsc{\tiny wp}}
|\cdot| U_{\textsc{\tiny wp}} | \|_\infty$.  $\| N_{\textsc{\tiny wp}} \|_\infty$ and $\| U_{\textsc{\tiny wp}} \|_\infty$
being good indicators of how large this first quantity might be.
We observe that the growth in $U_{\textsc{\tiny wp}}$
($\| U_{\textsc{\tiny wp}} \|_\infty$) is almost constant as we increase the number of tiles, 
unfortunately
the growth in $N_{\textsc{\tiny wp}}$
($\| N_{\textsc{\tiny wp}} \|_\infty$) is increasing quite significantly with $p$.
We note however that 
$\| |N_{\textsc{\tiny wp}} |\cdot| U_{\textsc{\tiny wp}} | \|_\infty$ is significantly smaller than
$\| N_{\textsc{\tiny wp}} \|_\infty \| U_{\textsc{\tiny wp}} \|_\infty$ which means that, hopefully, all the growth observed in $N$ 
does not end up in the error in the factorization.
We acknowledge that the mechanism behind this observation is not yet understood.

We report a last experiment that is worth noting. Since we are working with random
matrices, a reasonable pivoting strategy to consider is
\textit{Gaussian elimination with no pivoting} (GENP).
In this context, we would hope that GETWP is at least better than GENP.
It turns out that this is not the case for the backward error for
the factorization. We report for GENP a mean error of $2\cdot10^{-11}$
while it is the mean error is $7\cdot10^{-11}$ for GETWP and $p=128$.
Once more, we acknowledge that the mechanism behind this observation is not yet
understood.

\paragraph{Matrix Market matrices.}

\begin{figure}
\begin{minipage}{0.45 \textwidth}
\includegraphics[width=1.00 \textwidth]{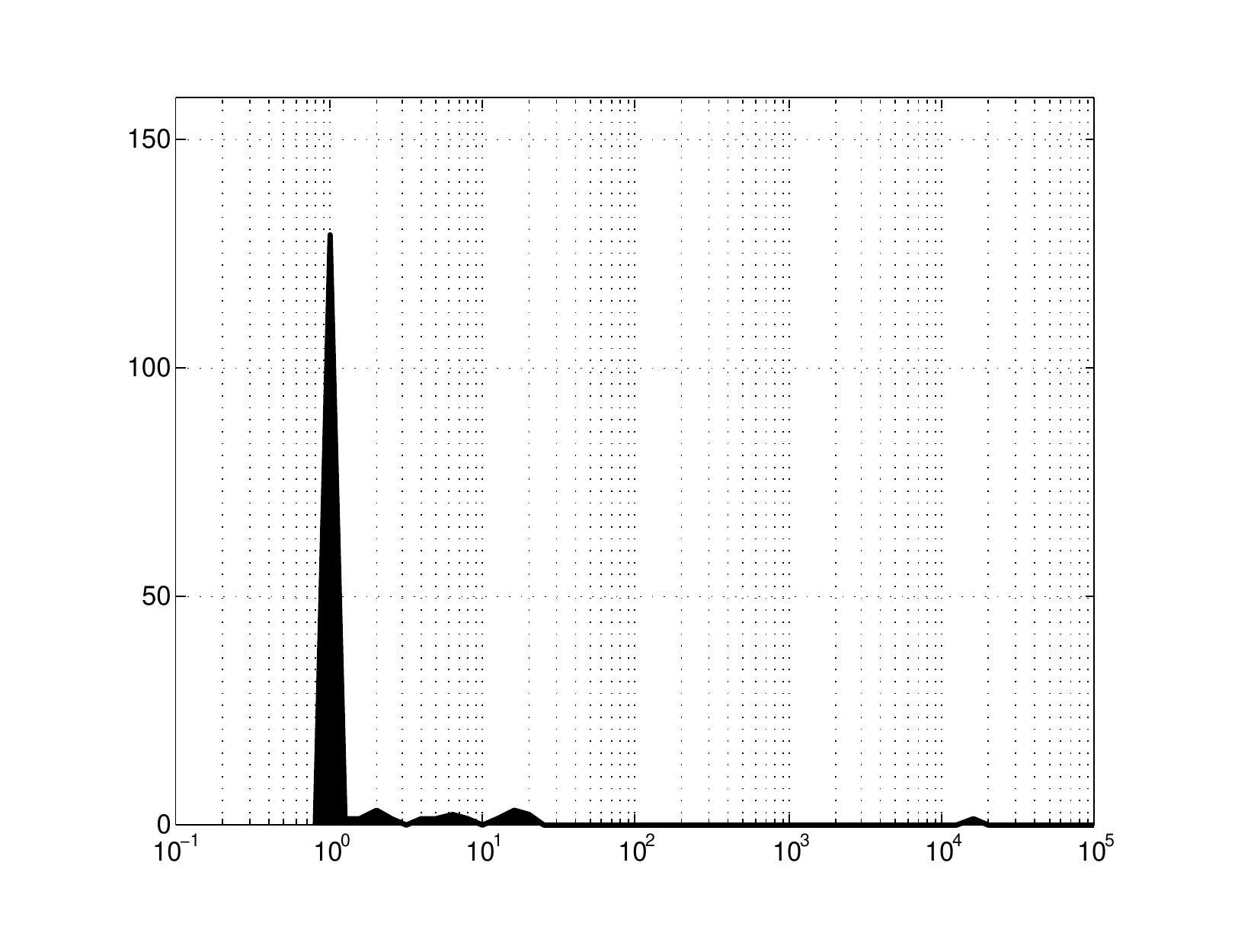}
\end{minipage}
\begin{minipage}{0.45 \textwidth}
\includegraphics[width=1.00 \textwidth]{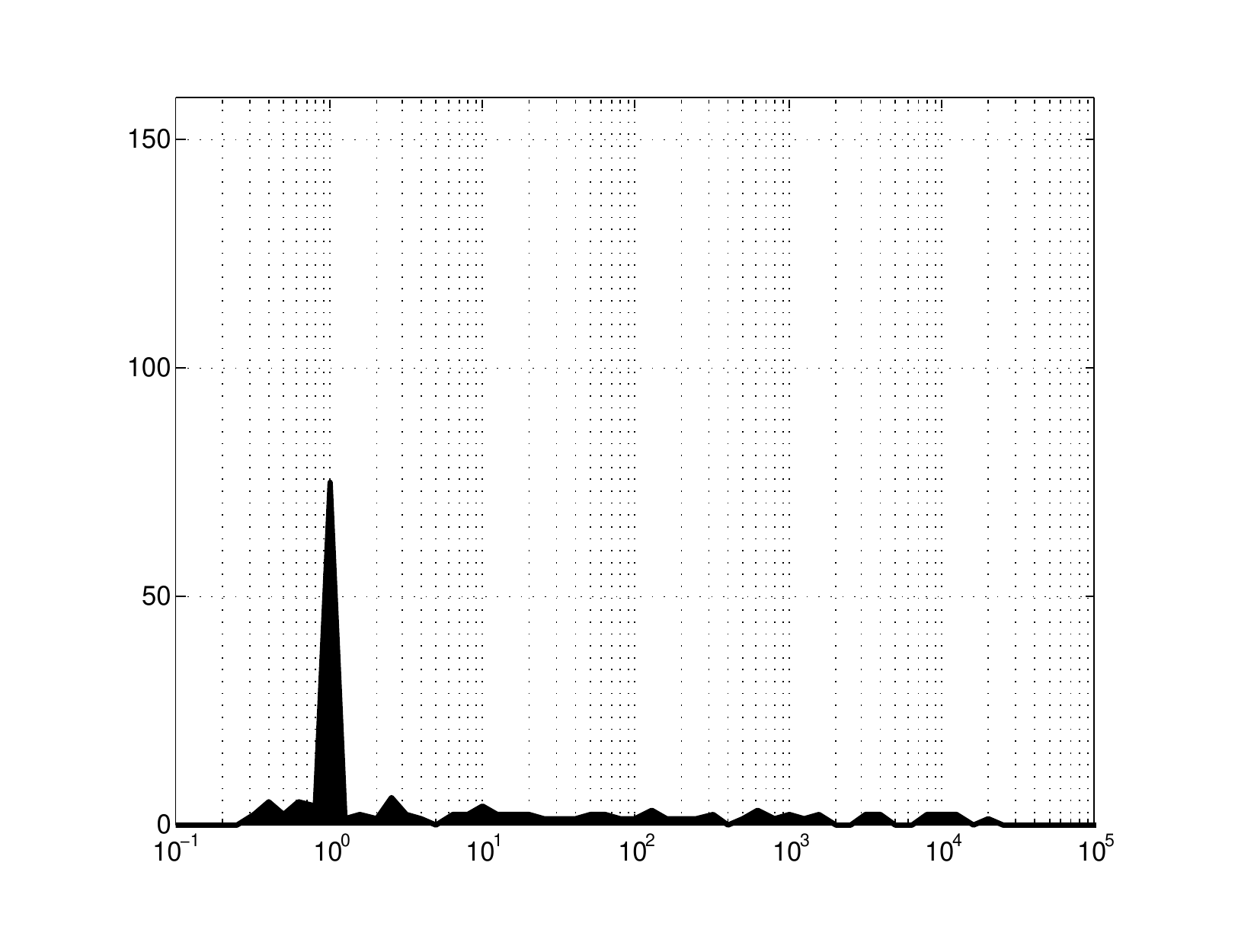}
\end{minipage}
\caption{\label{fig:market_res_003}
Histogram representing the distribution of the ratio $\| y - A x_{\textsc{\tiny wp}} \|_\infty/(\| A \|_\infty \|x_{\textsc{\tiny wp}}\|_\infty )$
for GETWP over  $\| y - A x_{\textsc{\tiny pp}} \|_\infty/(\| A \|_\infty \|x_{\textsc{\tiny pp}}\|_\infty )$ for GEPP (left) and the ratio of
$\| A - N_{\textsc{\tiny wp}} U_{\textsc{\tiny wp}} \|_\infty/\| A \|_\infty $ for GETWP over 
$\| PA - L_{\textsc{\tiny pp}} U_{\textsc{\tiny pp}} \|_\infty/\| A \|_\infty $ for GEPP (right).
Matrices are taken from the Matrix-Market collection~\cite{matrixmarket}
and right-hand sides are random.
}
\end{figure}

In Figure~\ref{fig:market_res_003}, we present stability results for GETWP
compared to GEPP on matrices from Matrix Market~\cite{matrixmarket}. At the
date of May 2008, we took all the matrices from Matrix Market with size ($n$)
between 1 and 6000 which are square, which are associated with \textit{Linear
System}, and which are in Matrix Market format (.mtx.gz)\footnote{This last restriction
implies  that we have discarded the five matrices that were in Harwell Boeing format (.pse.gz).} This
methodology provides us $159$ matrices.  For all these matrices, we assign a random
right-hand side $y$ (\texttt{y = randn(n,1);}) whether or not the matrix had a
prescribed right-hand side on Matrix Market. The tile size $b$ is a function of the 
matrix size $n$. In this experiment, we want to keep $p= n/b$ constant with $p=32$.

We note that some of these matrices will provide us with $U$ factors that have exact $0$'s
on their diagonals. This will result in \texttt{NaN} or \texttt{Inf} results.

On the left in Figure~\ref{fig:market_res_003}, we give the histogram of the 
ratio of the backward error for the solution when solving a linear system of equations
$$
\left(\frac{ \| y - A x_{\textsc{\tiny wp}} \|_\infty}{\| A \|_\infty \|x_{\textsc{\tiny wp}}\|_\infty }\right)
/
\left(\frac{ \| y - A x_{\textsc{\tiny pp}} \|_\infty}{\| A \|_\infty \|x_{\textsc{\tiny pp}}\|_\infty }\right).
$$
On the right in Figure~\ref{fig:market_res_003}, we give the histogram of the
ratio of the backward error for the factorization
$$
\left(\frac{ \| A - N_{\textsc{\tiny wp}}U_{\textsc{\tiny wp}} \|_\infty}{\| A \|_\infty }\right)
/
\left(\frac{ \| PA - L_{\textsc{\tiny pp}}U_{\textsc{\tiny pp}} \|_\infty}{\| A \|_\infty }\right).
$$

We have set any backward error (for the solution
or for the factorization) smaller than the machine precision 
at the level of the machine precision.

For 146 matrices out of 147\footnote{12 matrices
are indeed structurally singular and produce a 0 on the diagonal of the $U$ factor for both GEPP and GETWP},
we solve the linear system with a backward error for the solution lower than the one of
GEPP times $25$ (Figure~\ref{fig:market_res_003} left).

For 121 matrices out of 147,
we obtain a backward error for the factorization lower than the one of
GEPP times $25$ (Figure~\ref{fig:market_res_003} right).

The worst case matrix is in both case the matrix named \texttt{orani}.
Its condition number is about $10^4$ and its order is $n=2,529$. The ratio 
of backward error is 
$4.6\cdot10^7$ for the factorization and
$1.9\cdot10^4$ for the solution.
If we compare the norm of the factors, we get:
$$
\| N_{\textsc{\tiny wp}} \|_\infty = 2\cdot 10^5, \quad
\| U_{\textsc{\tiny wp}} \|_\infty = 6\cdot 10^4, \quad\textmd{and}\quad
\| L_{\textsc{\tiny pp}} \|_\infty = 20, \quad
\| U_{\textsc{\tiny pp}} \|_\infty = 10, \quad
$$
where we initially had $\| A \|_\infty = 9$.
We see that, for this special case, GETWP suffers of 
growth in the $N$ factor and growth in the  $U$ factor.

\section{Graph driven asynchronous execution}
\label{sec:async}
\begin{figure}[!h]
  \begin{center}
    \includegraphics[width=0.8\textwidth]{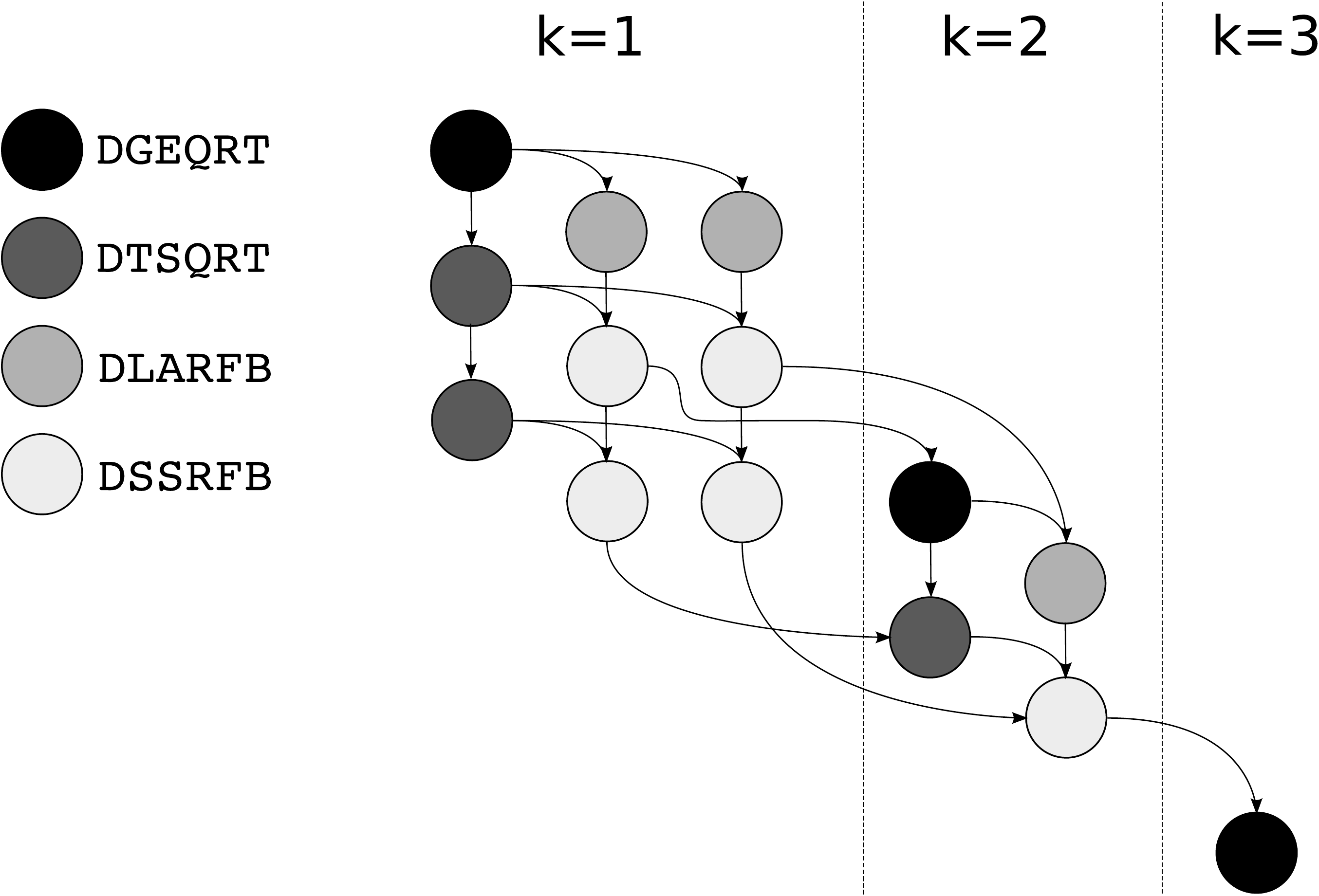}
  \caption{\label{fig:qr_dag}The dependency graph of
    Algorithm~\ref{alg:blkqr} on a matrix with $p=q=3$.}
  \end{center}
\end{figure}

Following the approach presented
in~\cite{Kurzak:2006:ILA,para06,tiledqr},
Algorithms~\ref{alg:blkchol},~\ref{alg:blkqr} and~\ref{alg:blklu} can
be represented as a Directed Acyclic Graph (DAG) where nodes are
computational tasks performed in kernel subroutines and where edges
represent the dependencies among them.  Figure~\ref{fig:qr_dag} show
the DAG for the tiled QR factorization when Algorithm~\ref{alg:blkqr}
is executed on a matrix with $p=q=3$.  Note that these DAGs have a
recursive structure and, thus, if $p_1 \ge p_2$ and $q_1 \ge q_2$ then
the DAG for a matrix of size $p_2 \times q_2$ is a subgraph of the DAG
for a matrix of size $p_1 \times q_1$. This property also holds for
most of the algorithms in LAPACK.

Once the DAG is known, the tasks can be scheduled asynchronously and
independently as long as the dependencies are not violated.  A
critical path can be identified in the DAG as the path that connects
all the nodes that have the higher number of outgoing edges; this non
conventional definition of critical path stems from the observation
that anticipating the execution of nodes with an higher number of
outgoing edges maximises the number of tasks in a ``ready''
state.
Based on this observation, a scheduling policy can be used, where
higher priority is assigned to those nodes that lie on the critical
path.  Clearly, in the case of our block algorithm for QR
factorization, the nodes associated to the \texttt{DGEQRT} subroutine
have the highest priority and then three other priority levels can be
defined for \texttt{DTSQRT}, \texttt{DLARFB} and \texttt{DSSRFB} in
descending order.

\begin{figure}[!h]
  \begin{center}
    \includegraphics[width=\textwidth]{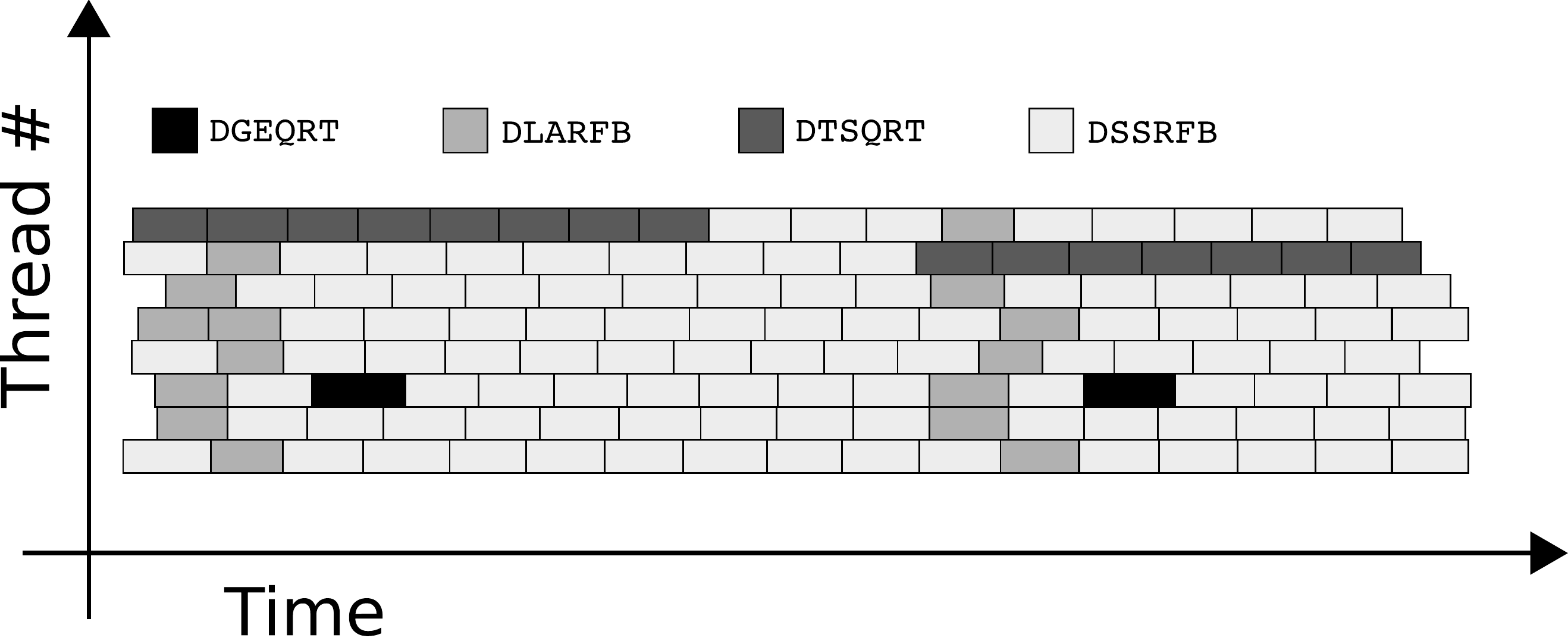}
  \caption{\label{fig:flow}The execution flow for dynamic scheduling,
    out of order execution of Algorithm~\ref{alg:blkqr}.}
  \end{center}
\end{figure}

This dynamic scheduling results in an out of order execution where
idle time is almost completely eliminated since only very loose
synchronization is required between the threads.
Figure~\ref{fig:flow} shows part of the execution flow of
Algorithm~\ref{alg:blkqr} using 8 cores machine when tasks are
dynamically scheduled based on dependencies in the DAG. Each line in
the execution flow shows which tasks are performed by one of the
threads involved in the factorization.

Figure~\ref{fig:flow} shows that all the idle times, which represent
the major scalability limit of the fork-join approach, can be removed
thanks to the very low synchronization requirements of the graph
driven execution. The graph driven execution also provides some degree
of adaptivity since tasks are scheduled to threads depending on the
availability of execution units.

\subsection{Implementation Details}
\label{sec:implementation}
The approach based on the combination of tiled algorithms, Block Data
Layout and graph driven, dynamic execution (as described,
respectively, in Sections~\ref{sec:tiled} and~\ref{sec:async}) has
been validated in a software implementation based on the pThreads
POSIX standard. The graph of dependencies is implicitly represented in
a shared progress table. Each thread in the pool is self-scheduled: it
check the shared progress table to identify a set of doable tasks and
then picks one of them according to a priority policy. Once a thread
terminates the execution of a task, it updates the progress table
accordingly. Because the centralized progress table may represent a
bottleneck and may imply more synchronization as the degree of
parallelism grows, the object of future work will be to distribute the
handling of the dependency graph.

Despite the choice of using the pThreads standard, the presented
approach may also be implemented by means of other technologies like,
for examples OpenMP or MPI or even an hybrid combination of them.

\section{Performance Results}
\label{sec:perf}

The performance of the tiled algorithms for Cholesky, QR ad LU
factorizations with dynamic scheduling of tasks (using ACML-4.0.0 BLAS
for tile computations) has been measured on the system described in
Table~\ref{tab:archs} and compared to the performance of the MKL-9.1
and ACML-4.0.0 implementations and to the fork-join approach, i.e.,
the standard algorithm for block factorizations of LAPACK associated
with multithreaded BLAS (ACML-4.0.0)\footnote{For both tiled
  algorithms and LAPACK, the choice of the underlying BLAS library is
  such that the highest performance possible is achieved}.
In the following figures, the tiled algorithms with dynamic scheduling
are referred to as PLASMA (Parallel Linear Algebra for Scalable
Multicore Architectures), the name of the project inside which the
presented work was developed.

\begin{table}[!h]
  \centering
  \begin{tabular}{p{1.4in}p{1.9in}}
\hline
                  & 8-way dual Opteron  \\
\hline
Architecture      &   AMD\textregistered  Opteron\textregistered  8214  \\
Clock speed       &   2.2 GHz              \\
\# cores          &   $2\times8=16$          \\
Peak performance  &   70.4 Gflop/s         \\
Memory            &   65 GB                 \\
Compiler suite    &   Intel 9.1             \\
BLAS libraries    &   MKL-9.1.023,  ACML-4.0.0\\
DGEMM performance &   57.5 Gflop/s\\
\hline
  \end{tabular}
  \caption{\label{tab:archs}Details of the system used for the following performance results.}
\end{table}

\begin{figure*}
\begin{minipage}[tl]{0.5\textwidth}
\begin{center}
{\includegraphics[width=1\textwidth]{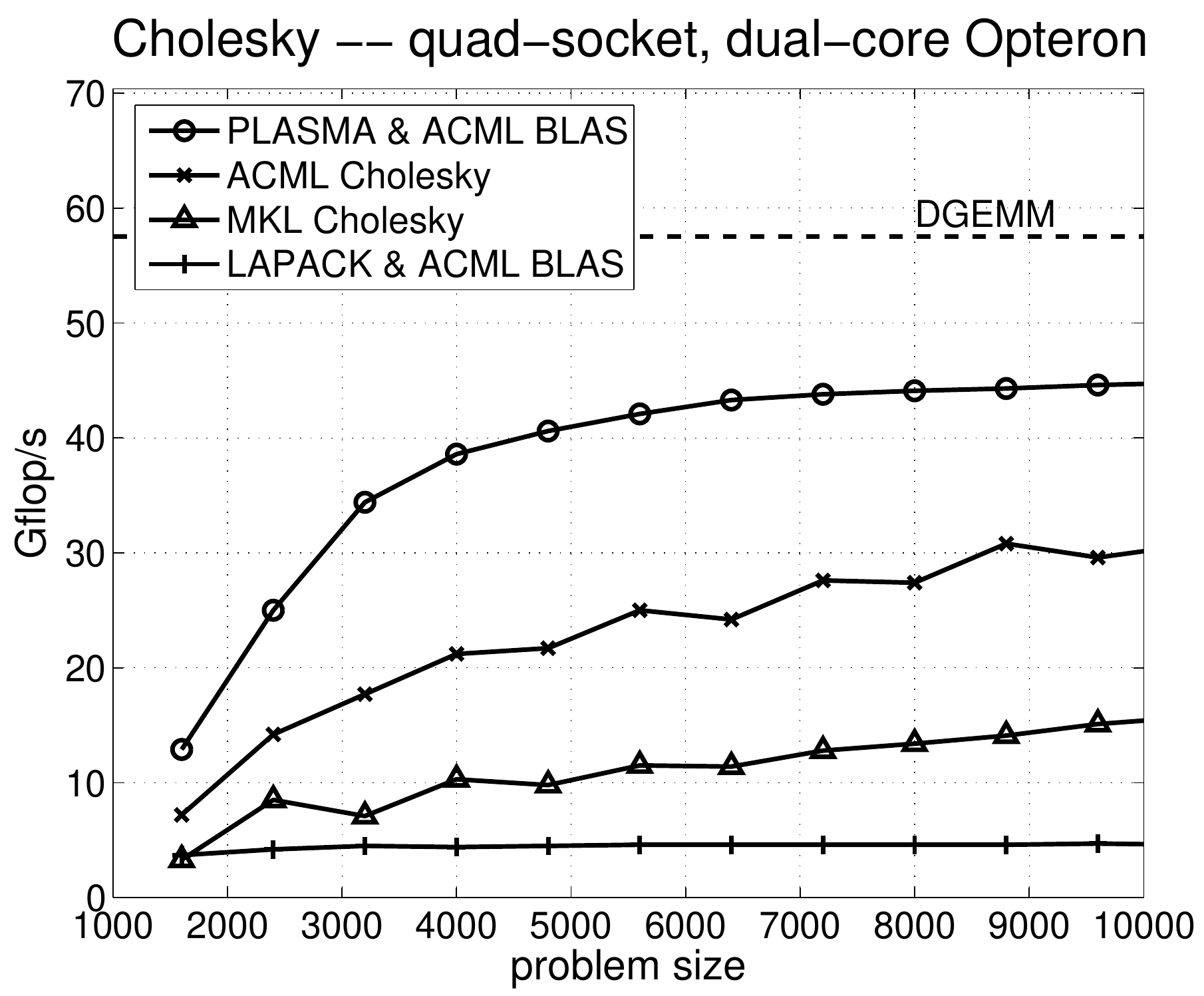}}
\caption{\label{fig:chol} Cholesky factorization.}
\end{center}
\end{minipage}
\hspace{0.25cm}
\begin{minipage}[tr]{0.5\textwidth}
\begin{center}
{\includegraphics[width=1\textwidth]{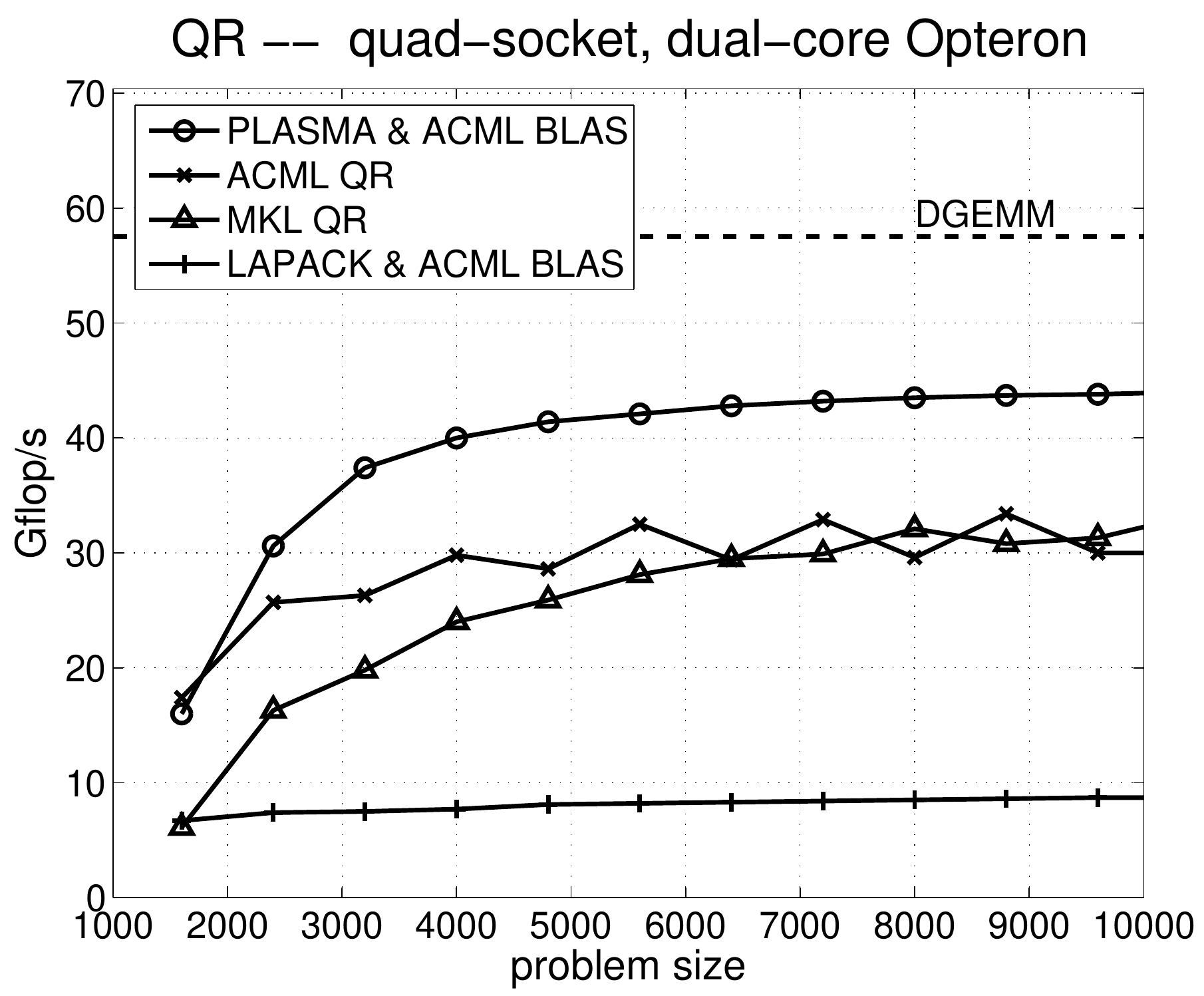}}
\caption{\label{fig:qr} QR factorization.}
\end{center}
\end{minipage}
\end{figure*}

\begin{figure*}
\begin{minipage}[tl]{0.5\textwidth}
\begin{center}
{\includegraphics[width=1\textwidth]{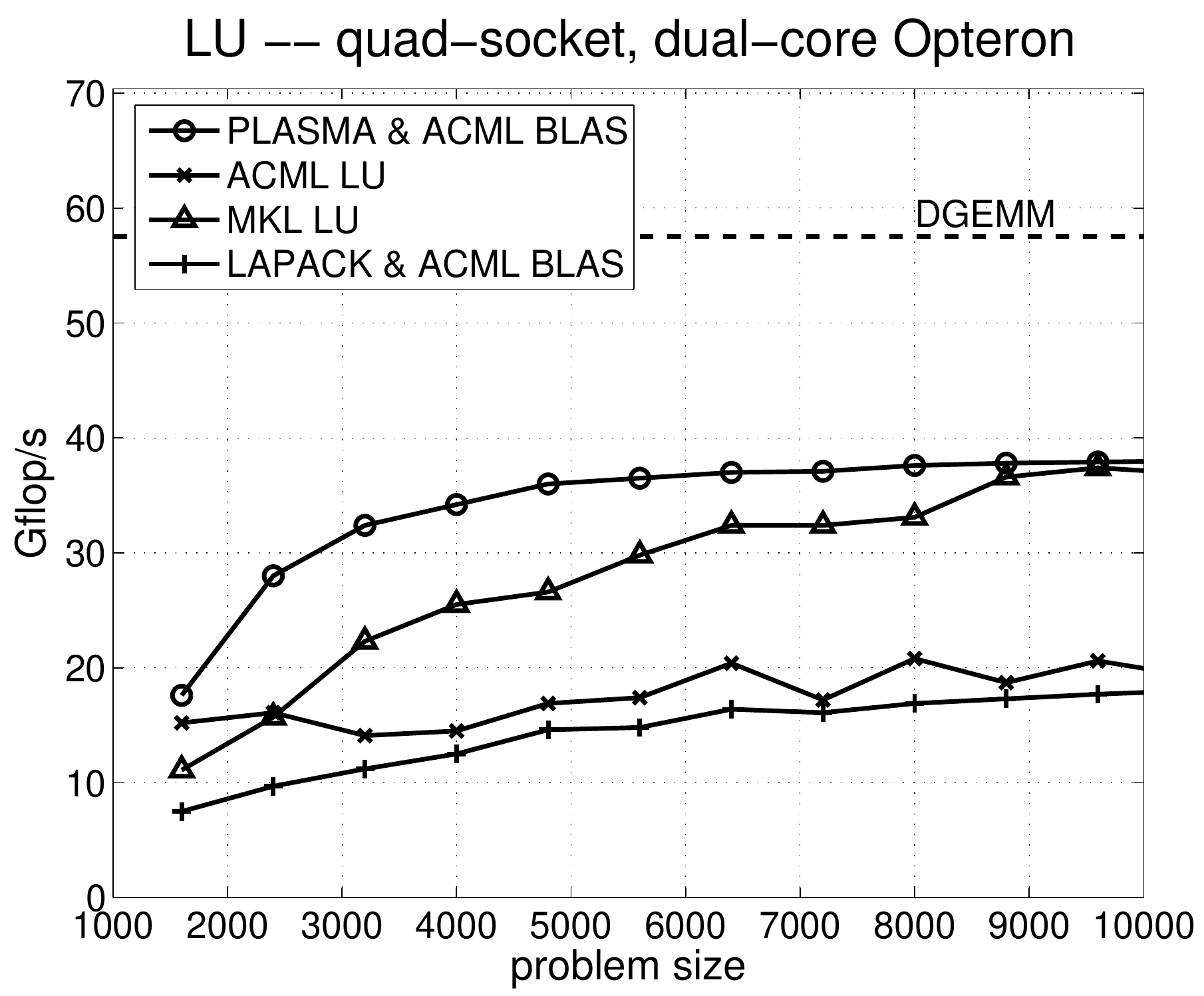}}
\caption{\label{fig:lu} LU factorization.}
\end{center}
\end{minipage}
\hspace{0.25cm}
\begin{minipage}[tr]{0.5\textwidth}
\begin{center}
{\includegraphics[width=1\textwidth]{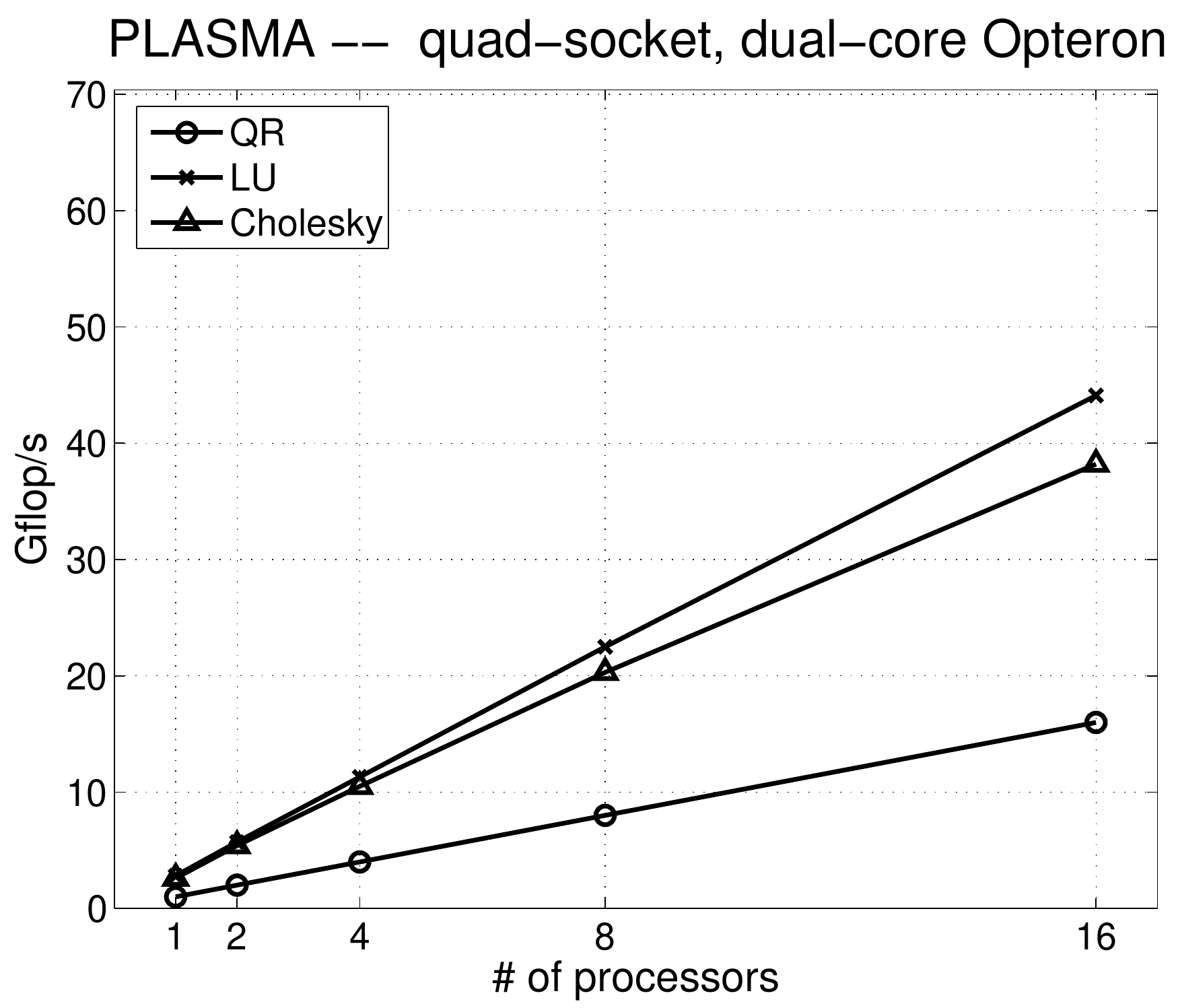}}
\caption{\label{fig:scal} PLASMA software: scalability.}
\end{center}
\end{minipage}
\end{figure*}

Figures~\ref{fig:chol},~\ref{fig:qr},~\ref{fig:lu} report the
performance of the Cholesky, QR and LU factorizations for the tiled
algorithms with dynamic scheduling, the MKL-9.1 and ACML-4.0.0
implementation and the LAPACK block algorithms with multithreaded
BLAS. For the tiled algorithms, the tile size and (for QR and LU) the
internal blocking size have been chosen in order to achieve the best
performance possible. As a reference, the tile size is in the range of
200 and the internal blocking size in the range of 20-40.  In the case
of the LAPACK block algorithms, the block size \footnote{the block
  size in the LAPACK algorithm sets the width of the panel.} has been
tuned in order to achieve the best performance; specifically the block
size was set to $100$. These graphs show the performance
measured using the maximum number of cores available on the system
(i.e., 16) with respect to the problem size. The axis of ordinates
has been scaled to reflect the theoretical peak performance of the
system (i.e. the top value is 70.4 Gflop/s) and, also, as a
reference, the performance of the matrix-matrix multiply (DGEMM) has
been reported.

Figure~\ref{fig:scal} shows the weak scalability, i.e. the flop rates versus the
number of cores when the local problem size is kept constant
(nloc=5,000) as the number of cores increases. 

In order to reflect the time to completion, in all the figures the
operation count of the tiled algorithms for QR and LU factorizations
is assumed to be the same as that of the LAPACK block algorithm; for
what discussed in Section~\ref{sec:cost}, this assumption is only
slightly inaccurate since the amount of extra flops can be considered
negligible for a correct choice of the internal blocking size $s$.

Figures~\ref{fig:chol} and~\ref{fig:qr} provide roughly the same
information:
the tiled algorithm combined with asynchronous graph driven execution
delivers higher execution rates than the fork-join approach
(i.e. LAPACK block algorithm with multithreaded BLAS) and performs
around 50\% better than a
vendor implementation of the operation. An important remark has to be
made for the Cholesky factorization: the {\it left-looking} variant
(see~\cite{552704} for more details) of the block algorithm is
implemented in LAPACK. This variant delivers very poor performance
when compared to the {\it right-looking} one; a sequential {\it
  right-looking} implementation of the Cholesky factorization that
uses multithreaded BLAS would run at higher speed than that measured on
the LAPACK version.

In the case of the LU factorization, even if it still provides a
considerable speedup with respect to the fork-join approach, the tiled
algorithm delivers, asymptotically, roughly the same performance as
the MKL-9.1 vendor implementation. This is mostly due to two main
reasons:
\begin{enumerate}
\item pivoting: in the block LAPACK algorithm, entire rows are swapped
  at once and, at most, $n$ swaps have to be performed where $n$ is
  the size of the problem. With pairwise pivoting, which is the
  pivoting scheme adopted in the tiled algorithm, at most $n^2/(2b)$
  can happen and all the swaps are performed in a very inefficient way
  since rows are swapped in pieces of size $b$.
\item internal blocking size: as shown in Section~\ref{sec:cost}, the
  flop count of the tiled algorithm grows by a factor of
  $1+s/(2b)$. To keep this extra cost limited to a negligible amount,
  a very small internal block size $s$ has to be chosen. This results
  in a performance loss due to the limitations of BLAS libraries on
  small size data.
\end{enumerate}

It must be noted, however, that the tiled algorithm for LU
factorization, reaches the asymptotic performance faster thus
providing considerable performance benefit for lower size
problems. This is a consequence of the fact that, once the values for
the tile size $b$ and blocking factor $s$ are fixed, the performance
of the BLAS operations is constant. The dynamic execution model
reduces the overhead of parallelization yielding the relatively steep
growth for the curve related to the tiled algorithm.

\section{Conclusions}
\label{sec:conclusions}

Even if a definition of multicore processor is still lacking, with
some speculation it is possible to define a limited set of
characteristics that a software should have in order to efficiently
take advantage of multiple execution units on a chip. 

The work presented here follows a path established by the same authors
in~\cite{cell_chol,Kurzak:2006:ILA,para06,tiledqr} exploiting and
reinterpreting ideas already studied in the
past~\cite{1014508,670985,76287,essl01}.

The discussed approach suggests that fine granularity and asynchronous
execution models are desirable properties in order to achieve high
performance on multicore architectures due to high degrees of
parallelism, increased importance of local data reuse and the
necessity to hide the latency of access to memory.

Performance results presented in Section~\ref{sec:perf} support this
reasoning by showing how the usage of fine granularity, tiled
algorithms together with a graph driven, asynchronous execution model
can provide considerable benefits over the traditional fork-join
approach and also vendor implementations.

The quality of the discussed approach is also supported by results
achieved by the FLAME group at University of Texas
Austin~\cite{1248397,vdgqr}.

\bibliography{tiled_algs}  

\begin{thebibliography}{10}

\bibitem{top500}
\url{http://top500.org}.

\bibitem{mkl}
\url{http://www.intel.com/cd/software/products/asmo-na/eng/307757.htm}.

\bibitem{matrixmarket}
\url{http://math.nist.gov/MatrixMarket/}.

\bibitem{polaris}
Teraflops research chip.
\newblock \url{http://www.intel.com/research/platform/terascale/teraflops.htm}.

\bibitem{essl01}
{IBM} engineering and scientific subroutine library for {AIX} version 3,
  release 3., December 2001.
\newblock {IBM} Pub. No. SA22-7272-04.

\bibitem{76287}
R.~C. Agarwal and F.~G. Gustavson.
\newblock Vector and parallel algorithms for {C}holesky factorization on {IBM}
  3090.
\newblock In {\em Supercomputing '89: Proceedings of the 1989 ACM/IEEE
  conference on Supercomputing}, pages 225--233, New York, NY, USA, 1989. ACM.

\bibitem{tiledqr}
{A}lfredo {B}uttari, {J}ulien {L}angou, {J}akub {K}urzak, and {J}ack
  {D}ongarra.
\newblock {P}arallel {T}iled {QR} {F}actorization for {M}ulticore
  {A}rchitectures.
\newblock Technical report, University of Tennessee Knoxville, 2007.
\newblock UT-CS-07-598. 23-July-2007.

\bibitem{lapack:99}
E.~Anderson, Z.~Bai, C.~Bischof, S.~Blackford, J.~Demmel, J.~Dongarra, J.~Du
  Croz, A.~Greenbaum, S.~Hammarling, A.~McKenney, and D.~Sorensen.
\newblock {\em {LAPACK} {U}sers' {G}uide}.
\newblock SIAM, Philadelphia, 3. edition, 1999.

\bibitem{210517}
M.~W. Berry, J.~J. Dongarra, and Y.~Kim.
\newblock A parallel algorithm for the reduction of a nonsymmetric matrix to
  block upper-{H}essenberg form.
\newblock {\em Parallel Comput.}, 21(8):1189--1211, 1995.

\bibitem{para06}
A.~Buttari, J.~Dongarra, J.~Kurzak, J.~Langou, P.~Luszczek, and S.~Tomov.
\newblock The impact of multicore on {M}ath software.
\newblock In B.~K\aa gstr\"{o}m~et al., editor, {\em State of the Art in
  Scientific Computing. 8th International Workshop, PARA 2006, Umea, Sweden,
  June 18-21, 2006}, volume 4699 of {\em Lecture Notes in Computer Science},
  pages 1--10, 2007.

\bibitem{1248397}
E.~Chan, E.~S. Quintana-Ort{\'{\i}}, G.~Quintana-Ort\'{\i}, and R.~van~de
  Geijn.
\newblock Supermatrix out-of-order scheduling of matrix operations for {SMP}
  and multi-core architectures.
\newblock In {\em SPAA '07: Proceedings of the nineteenth annual ACM symposium
  on Parallel algorithms and architectures}, pages 116--125, New York, NY, USA,
  2007. ACM Press.

\bibitem{scalapack:96}
J.~Choi, J.~Demmel, I.~Dhillon, J.~Dongarra, S.~Ostrouchov, A.~Petitet,
  K.~Stanley, D.~Walker, and R.~C. Whaley.
\newblock {ScaLAPACK}: A portable linear algebra library for distributed memory
  computers - design issues and performance.
\newblock {\em Computer Physics Communications}, 97:1--15, 1996.
\newblock (also as LAPACK Working Note \#95).

\bibitem{666023}
J.~Choi, J.~Dongarra, S.~Ostrouchov, A.~Petitet, D.~W. Walker, and R.~C.
  Whaley.
\newblock A proposal for a set of parallel basic linear algebra subprograms.
\newblock In {\em PARA '95: Proceedings of the Second International Workshop on
  Applied Parallel Computing, Computations in Physics, Chemistry and
  Engineering Science}, pages 107--114, London, UK, 1996. Springer-Verlag.

\bibitem{hep}
J.~J. Dongarra and R.~E. Hiromoto.
\newblock A collection of parallel linear equations routines for the {Denelcor
  HEP}.
\newblock {\em {P}arallel {C}omputing}, 1(2):133--142, December 1984.

\bibitem{552704}
Jack~J. Dongarra, Iain~S. Duff, Danny~C. Sorensen, and Henk~A. {van der Vorst}.
\newblock {\em Numerical Linear Algebra for High Performance Computers}.
\newblock Society for Industrial and Applied Mathematics, Philadelphia, PA,
  USA, 1998.

\bibitem{kag-gus}
Erik Elmroth, Fred Gustavson, Isak Jonsson, and Bo~K\aa gstr\"{o}m.
\newblock Recursive blocked algorithms and hybrid data structures for dense
  matrix library software.
\newblock {\em SIAM Review}, 46(1):3--45, 2004.

\bibitem{gaps:90}
K.~A. Gallivan, R.~J. Plemmons, and A.~H. Sameh.
\newblock Parallel algorithms for dense linear algebra computations.
\newblock {\em SIAM Review}, 32(1):54--135, March 1990.

\bibitem{golubvanloan}
G.~Golub and C.~Van~Loan.
\newblock {\em Matrix Computations}.
\newblock Johns Hopkins University Press, Baltimore, MD, 3rd edition, 1996.

\bibitem{gotoblas}
K.~Goto and R.~van~de Geijn.
\newblock High-performance implementation of the level-3 {BLAS}.
\newblock Technical Report TR-2006-23, The University of Texas at Austin,
  Department of Computer Sciences., 2006.
\newblock FLAME Working Note 20.

\bibitem{vdgqr}
{G}regorio~{Q}uintana {O}rti, {E}nrique~{Q}uintana {O}rti, {E}rnie {C}han
  {F}ield {G}.~{V}an {Z}ee, and {R}obert van~de {G}eijn.
\newblock {S}cheduling of {QR} {F}actorization {A}lgorithms on {SMP} and
  {M}ulti-{C}ore {A}rchitectures.
\newblock Technical report, The University of Texas at Austin, Department of
  Computer Sciences, 2007.
\newblock Flame Working Note 24. 31-July-2007.

\bibitem{1055534}
B.~C. Gunter and R.~A. van~de Geijn.
\newblock Parallel out-of-core computation and updating of the {QR}
  factorization.
\newblock {\em ACM Trans. Math. Softw.}, 31(1):60--78, 2005.

\bibitem{three}
F.~Gustavson, L.~Karlsson, and B.~K\aa gstr\"{o}m.
\newblock Three algorithms for {C}holesky factorization on distributed memory
  using packed storage.
\newblock In B.~K\aa gstr\"{o}m~et al., editor, {\em State of the Art in
  Scientific Computing. 8th International Workshop, PARA 2006, Umea, Sweden,
  June 18-21, 2006}, volume 4699 of {\em Lecture Notes in Computer Science},
  pages 550--559, 2007.

\bibitem{279535}
F.~G. Gustavson.
\newblock Recursion leads to automatic variable blocking for dense
  linear-algebra algorithms.
\newblock {\em IBM J. Res. Dev.}, 41(6):737--756, 1997.

\bibitem{670985}
F.~G. Gustavson.
\newblock New generalized data structures for matrices lead to a variety of
  high performance algorithms.
\newblock In {\em PPAM '01: Proceedings of the th International Conference on
  Parallel Processing and Applied Mathematics-Revised Papers}, pages 418--436,
  London, UK, 2002. Springer-Verlag.

\bibitem{1014508}
F.~G. Gustavson.
\newblock High-performance linear algebra algorithms using new generalized data
  structures for matrices.
\newblock {\em IBM J. Res. Dev.}, 47(1):31--55, 2003.

\bibitem{hihi:89}
N.~J. Higham and D.~J. Higham.
\newblock Large growth factors in {Gaussian} elimination with pivoting.
\newblock {\em SIAM J. Matrix Anal. Appl.}, 10(2):155--164, April 1989.

\bibitem{DBLP:conf/para/JoffrainQG04}
Thierry Joffrain, Enrique~S. Quintana-Ort\'{\i}, and Robert~A. van~de Geijn.
\newblock Rapid development of high-performance out-of-core solvers.
\newblock In Jack Dongarra, Kaj Madsen, and Jerzy Wasniewski, editors, {\em
  PARA}, volume 3732 of {\em Lecture Notes in Computer Science}, pages
  413--422. Springer, 2004.

\bibitem{cell_chol}
J.~Kurzak, A.~Buttari, and J.~Dongarra.
\newblock Solving systems of linear equations on the {CELL} processor using
  {C}holesky factorization.
\newblock Technical Report UT-CS-07-596, Innovative Computing Laboratory,
  University of Tennessee Knoxville, April 2007.

\bibitem{Kurzak:2006:ILA}
J.~Kurzak and J.~Dongarra.
\newblock Implementing linear algebra routines on multi-core processors with
  pipelining and a look ahead.
\newblock LAPACK Working Note 178, September 2006.
\newblock Also available as UT-CS-06-581.

\bibitem{jordan}
R.~E. Lord, J.~S. Kowalik, and S.~P. Kumar.
\newblock Solving linear algebraic equations on an {MIMD} computer.
\newblock {\em J. ACM}, 30(1):103--117, 1983.

\bibitem{isscc_2005_cell_desing}
D.~{Pham}, S.~{Asano}, M.~{Bolliger}, M.~N. {Day}, H.~P. {Hofstee}, C.~{Johns},
  J.~{Kahle}, A.~{Kameyama}, J.~{Keaty}, Y.~{Masubuchi}, M.~{Riley},
  D.~{Shippy}, D.~{Stasiak}, M.~{Suzuoki}, M.~{Wang}, J.~{Warnock},
  S.~{Weitzel}, D.~{Wendel}, T.~{Yamazaki}, and K.~{Yazawa}.
\newblock The design and implementation of a first-generation {CELL} processor.
\newblock In {\em IEEE International Solid-State Circuits Conference}, pages
  184--185, 2005.

\bibitem{vdgooclu}
E.~Quintana-Ort\'{\i} and R.~van~de Geijn.
\newblock Updating an {LU} factorization with pivoting.
\newblock Technical Report TR-2006-42, The University of Texas at Austin,
  Department of Computer Sciences, 2006.
\newblock FLAME Working Note 21.

\bibitem{reid:71}
J.~K. Reid.
\newblock A note on the stability of {G}aussian elimination.
\newblock {\em IMA Journal of Applied Mathematics}, 8(3):374--375, 1971.

\bibitem{64889}
R.~Schreiber and C.~van Loan.
\newblock A storage-efficient {WY} representation for products of {H}ouseholder
  transformations.
\newblock {\em SIAM J. Sci. Stat. Comput.}, 10(1):53--57, 1989.

\bibitem{sore:85}
D.~C. Sorensen.
\newblock Analysis of pairwise pivoting in {Gaussian} elimination.
\newblock {\em IEEE Trans. Comput.}, 34(3):274--278, 1985.

\bibitem{stew:98}
G.~W. Stewart.
\newblock {\em Matrix Algorithms}, volume~1.
\newblock SIAM, Philadelphia, 1. edition, 1998.

\bibitem{trsc:90}
L.~N. Trefethen and R.~S. Schreiber.
\newblock Average-case stability of {Gaussian} elimination.
\newblock {\em SIAM J. Matrix Anal. Appl.}, 11(3):335--480, July 1990.

\bibitem{ATLAS}
R.~C. Whaley, A.~Petitet, and J.~Dongarra.
\newblock Automated empirical optimization of software and the {ATLAS} project.
\newblock {\em Parallel Computing}, 27(1--2):3--25, 2001.

\bibitem{wilk:65}
J.~H. Wilkinson.
\newblock {\em The Algebraic Eigenvalue Problem}.
\newblock Oxford University Press, 1965.

\bibitem{yip_ooc}
E.~L. Yip.
\newblock Fortran subroutines for out-of-core solutions of large complex linear
  systems.
\newblock Technical Report CR-159142, NASA, November 1979.

\end{thebibliography}
\bibliographystyle{plain}     

\end{document}